\documentstyle[graphicx,fleqn,twoside,12pt,epsf]{article}
\providecommand{\LyX}{L\kern-.1667em\lower.25em\hbox{Y}\kern-.125emX\@}

\input{ueur.fd}
\DeclareMathVersion{euler}
\DeclareMathVersion{eulerbold}
\SetSymbolFont{letters}  {euler}{U}{eur} {m}{n}
\SetSymbolFont{letters}  {eulerbold}{U}{eur} {b}{n}
\SetSymbolFont{operators}{eulerbold}{OT1}{cmr} {bx}{n}
\SetSymbolFont{symbols}  {eulerbold}{OMS}{cmsy}{b}{n}
\SetMathAlphabet\mathsf{eulerbold}{OT1}{cmss}{bx}{n}
\SetMathAlphabet\mathit{eulerbold}{OT1}{cmr}{bx}{it}
\def\eulermath{\@nomath\eulermath
              \mathversion{euler}}
\def\uneulermath{\@nomath\uneulermath
              \mathversion{normal}}
\def\eulerbfmath{\@nomath\eulerbfmath
              \mathversion{eulerbold}}
\def\uneulerbfmath{\@nomath\uneulerbfmath
              \mathversion{normal}}

\let\newcommand=\providecommand

\makeatother

\begin{document}

\title{Numerical analysis of the one-mode solutions in the Fermi-Pasta-Ulam system.\date{}}
\author{A. Cafarella, M. Leo and R.A.\ Leo \\
Dipartimento di Fisica dell'Universit\`{a}, 73100 Lecce, Italy, \\}
\maketitle

\begin{abstract}
\noindent The stability of the one-mode nonlinear solutions of the Fermi-Pasta-Ulam - $\beta$ system is numerically
investigated. No external perturbation is considered for the one-mode exact analytical solutions, the only
perturbation being that introduced by computational errors in numerical integration of motion equations.
The threshold energy for the excitation of the other normal modes and the dynamics of this excitation are studied
as a function of the parameter $\mu$ characterizing the nonlinearity, the energy density $\epsilon$ and the number
$N$ of particles of the system. The achieved results confirm in part previous results, obtained with a linear
analysis of the problem of the stability, and clarify the dynamics by which the one-mode exchanges energy with
the other modes with increasing energy density. In a range of energy density near the threshold value and for
various values of the number of particles $N$, the nonlinear one-mode exchanges energy with the other linear modes
for a very short time, immediately recovering all its initial energy. This sort of recurrence is very similar to
Fermi recurrences, even if in the Fermi recurrences the energy of the initially excited mode changes continuously
and only periodically recovers its initial value. A tentative explanation of this intermittent behaviour, in
terms of Floquet's theorem, is proposed.
\end{abstract}
\noindent PACS numbers: 05.45.+b; 63.20.Ry; 63.10. + a \par
\noindent Keywords: Anharmonic lattices; Energy equipartition; Periodic
solutions; Stability
\vfill\eject

\section{ Introduction}

Since the computer experiment of Fermi-Pasta-Ulam (FPU) \cite{fermi}, many theoretical and
numerical investigations followed to explain the unexpected results of the
experiment \cite{ford} - \cite{zab}. What one expected according to a
theorem of Poincar\`e \cite{poi} and a theorem proved by Fermi \cite{fer}
himself in his youth, and what was instead observed, has been narrated  in
several papers (see e.g. \cite{fo}) in which various aspects of the
experiment have been analysed in the framework of the KAM theorem \cite{kam},
ergodic problem \cite{saito}, statistical mechanics and chaotic behaviour of dynamical systems
\cite{livi,casetti}.

It is well known \cite{kantz,vulpiani} that, for a periodic FPU-$\beta$ chain with an even number
of oscillators, an initial condition with only a set of excited modes, all having even (resp. odd)
indices only, cannot lead to the excitation of modes having odd (resp. even) indices.
This means that, if one considers the set of all modes partitioned in the two subsets of
the even and odd modes, an initial excitation, completely contained in one of the two subsets, cannot 
propagate to the other. 

There are also other partitions. For example, partitions exist for which a subset contains one mode only 
\cite{poggi}. More specifically, for each of the modes 
\begin{equation}
n = \frac{N}{4}; ~~\frac{N}{3}; ~~\frac{N}{2}; ~~\frac{2}{3}N; ~~\frac{3}{4}N,
\label{eq:nnnnn}
\end{equation}
\noindent (of course when $N$ has the divisibility property required for $n$ in (\ref{eq:nnnnn})
to be an integer) one has that, if  only one of these mode is initially excited, 
 it remains excited without transferring energy to any other mode.

An important problem is obviously the stability of these one mode solutions (OMS's), since it is reasonable to 
expect some relation between the loss of their stability and the onset of chaos in the system. In some sense, 
the destabilization of the simplest nonlinear modes can provide an "upper  bound" estimate for the onset 
of the large scale chaos. The analysis of the stability of a generic OMS is very difficult from a 
mathematical point of view. Only for the case $n = N/2$ there are analytical results ( see for example \cite{bud}, 
\cite{fla}, \cite{poggi} ) which estimate the threshold energy density for the mode to become unstable. In fact 
the mode $n = N/2$ is the simpler one, since the different components of the perturbation, in modal space, are all 
decoupled and are described  by a single Lam\'e equation. 

In this paper we numerically revisit the problem of stability of the OMS's and we present 
the results of a global analysis. We made a  numerical study of the stability of the OMS's 
as a function of the number $N$ of particles and of the product $\epsilon \mu$, where $\epsilon$ is the energy 
density and $\mu$ is the parameter of nonlinearity in the Hamiltonian of the system. The analysis is based on  
the numerical integration of the  full nonlinear FPU model. No external perturbation is considered 
for the OMS, the only perturbation being that introduced by computational errors in numerical 
integration of motion equations. This simple method works very well: the results obtained 
confirm previous results obtained with a linear analysis of the problem of the stability(\cite{bud}, \cite{fla}, 
\cite{poggi}) and clarify the dynamics with which a nonlinear one-mode exchanges energy with the other linear 
modes, with increasing energy densities. We have found that, in a large range of initial excitation energy density, 
the energy of the OMS's keeps constant for very long times, and for short times it is partially transferred to other 
linear modes; furthermore the OMS's corresponding to the two values $n = \frac{N}{4}$ and $n = \frac{3}{4}N$ and 
the OMS's corresponding to $n = \frac{N}{3}$ and $n = \frac{2}{3}N$ have the same stability 
properties.     

\section{The FPU system}

The FPU system is a one-dimensional chain of oscillators (with unit mass), with weakly nonlinear nearest-neighbour 
interaction. The nonlinear forces considered are quadratic, cubic and broken linear ones. In the present work we 
consider only the FPU-$\beta$ model (nonlinear cubic forces), with periodic boundary conditions. Calling $q_n$ 
and $p_n$ the coordinates and the momenta of the oscillators, the model is defined by the following Hamiltonian: 

\begin{equation}
H=H_0 + H_1,
\label{eq:h}
\end{equation}

being

\begin{equation}
H_0 = \frac{1}{2} \sum_{k=1}^{N}{p_k}^2 + \frac{1}{2} \sum_{k=1}^{N}(q_{k+
1} - q_k)^2,
\label{eq:h0}
\end{equation}

\begin{equation}
H_1 = \frac{\mu}{4} \sum_{k=1}^{N}(q_{k+1} - q_k)^4,
\label{eq:h1}
\end{equation}

\noindent with $q_{N+1} = q_1$.

\noindent If we introduce the  normal coordinates  $Q_k$ ~and~ $P_k$  of the normal modes through the relations: \par
\begin{equation}
Q_k = \sum_{j=1}^{N} S_{kj} q_j,
\label{eq:qk}
\end{equation}

\begin{equation}
P_k = \sum_{j=1}^{N} S_{kj} p_j,
\label{eq:pk}
\end{equation}

\begin{equation}
S_{kj} = \frac{1}{\sqrt{N}} (\sin{\frac{2 \pi kj}{N}} + \cos{\frac{2 \pi 
kj}{N}}),
\label{eq:skj}
\end{equation}

\noindent the harmonic energy of mode $k$ is:
\begin{equation} 
E_k = \frac{1}{2} ({P_k}^2 + {\omega_k}^2 {Q_k}^2)
\label{eq:el}
\end{equation}

\noindent where, in the case of periodic boundary conditions,

\begin{equation}
{\omega^{2}}_k = 4 \sin^{2}{\frac{\pi k}{N}}.
\label{eq:omeq}
\end{equation}

\noindent We have $\omega_k = \omega_{N-k}$, so that there are only 
$\frac{N}{2}$ different frequencies (if we assume $N$ even, for simplicity). \par
From (\ref{eq:h0} - \ref{eq:h1}), the Hamilton equations are obtained in the variables $q_k$ and 
$p_k$, which, integrated by standard methods, allow to calculate the normal 
modes and the energy of each mode. 

\section{The one-mode solutions}

Let us first consider the case $\mu = 0$. In this case all normal modes 
oscillate independently one from another and their energies $E_{k}$ are constants of motion. 

In the anharmonic case ($\mu \neq 0$), the normal modes are not independent 
and the variables $Q_k$ have not simple sinusoidal oscillations. All the
modes are coupled and the differential equation for the $k$-th mode is
 \cite{poggi}:

\begin{equation}
\stackrel{..}{Q_k} = F_k(Q_1,...,Q_{N-1}),~~~~~~    (k= 1,....,N-1),
\label{eq:qdp}
\end{equation}

\noindent where $F_k$ is the generalized force in normal coordinate space, given by:

\begin{equation}
F_k(Q_1,...,Q_{n-1}) = -{\omega_k}^2 Q_k - \frac{\mu\omega_k}{2N}\sum_{
i,j,l}^{N-1}\omega_i\omega_j\omega_l C_{kijl} Q_i Q_j Q_l
\label{eq:f}
\end{equation}

\noindent and

\begin{equation}
C_{ijkl} = -\triangle_{i+j+k+l} + \triangle_{i+j-k-l} + \triangle_
{i-j+k-l} + \triangle_{i-j-k+l},
\label{eq:c}
\end{equation}

\noindent being $ \triangle_{k} = (-1)^{m} $ for $k = m N$, if $m$ is a positive integer,  and                                                                                                                       
$\triangle_{k} = 0$ otherwise.

The nonlinear one-mode solutions correspond to the values of $n$: \par   
\begin{equation}
n = \frac{N}{4}; ~~\frac{N}{3}; ~~\frac{N}{2}; ~~\frac{2}{3}N; ~~\frac{3}{4}N.
\label{eq:oms}
\end{equation}

From (\ref{eq:f}) and (\ref{eq:c}), one deduces \cite{poggi} that, if  only one of these modes is initially 
excited, it remains excited without transferring energy to any other mode. In this case, the equation of 
motion for the excited mode amplitude $Q_n$ is:

\begin{equation}  
\stackrel{..}Q_n = -{\omega_n}^2~Q_n -\frac{\mu{\omega_n}^4C_{nnnn}}{2N}{Q_n}^3.
\label{eq:q2p}
\end{equation}

If we assume that at time $t=0$ ~$Q_n \neq 0$ and $P_{n} = 0$, the solution of (\ref{eq:q2p}) is:
\begin{equation}
Q_n( t ) = A~cn(\Omega_n t,k),
\label{eq:cn}
\end{equation}

\noindent where $\Omega_n$ and the modulus $k$ of the Jacobi elliptic function 
$cn$ both depend on A:

\begin{equation}
\Omega_n = \omega_n \sqrt{1+ \delta_n A^2},
\label{eq:omen}
\end{equation}

\begin{equation}
k = \sqrt{\frac{\delta_n A^2}{2(1+\delta_n A^2)}},
\label{eq:kappa}
\end{equation}

\noindent with $\delta_n = \mu~\epsilon C_{nnnn}/2N$. \par 
The solution (\ref{eq:cn}) is periodic  with period $T_n = 4 K(k)/\Omega_n$ where $K(k)$ is the complete elliptic 
integral of the first kind. The energy of the mode is:
\begin{equation}
E_{n} = \frac{1}{2} (P_{n}^{2} + \omega_{n}^{2}~ Q_{n}^{2} + \mu~ \frac{\omega_{n}^{4}~Q_{n}^{4} C_{nnnn}}{4~N}).
\label{eq:en}
\end{equation}
In the next section, the problem of the stability of the OMS corresponding to $n = N/2$ will be analysed. 
\bigskip
\section{The stability of the one-mode solution $N/2$}

The stability properties of the nonlinear mode $N/2$ was studied analytically some years ago. In 
\cite{bud}, the nonlinear mode $N/2$ was named "the out of phase mode". This expression derives from the fact 
that from (\ref{eq:qk}) and the properties of $S_{kj}$ one has, if the only excited mode is the mode $N/2$:
\bigskip
\begin{equation} 
q_{k}=\frac{1}{\sqrt{N}} (-1)^{k}~~Q_{N/2} 
\label{eq:k}
\end{equation}
\bigskip
and then: \par
\begin{equation}
q_{k+1}(t) = - q_{k}(t), ~~~~k = 1, 2, \ldots, N.
\label{eq:q-q}
\end{equation}
In \cite{bud}, the stability analysis of the out of phase mode (mode $N/2$) starts from the equations of motion for 
the variables $q_{k}$. Due to the relations (\ref{eq:q-q}), these equations reduce to a single equation, describing 
the anharmonic oscillations of each particle, whose solution is the Jacobi elliptic cosine function. Perturbing this 
solution, and passing to normal modes variables $Q_{k}$ one obtains a Lam\'e equation. The stability of the 
solutions of this equation, which is an example of Hill's equation, and then the stability of the mode $N/2$, 
is studied with the Floquet theory. A numerical analysis shows that the first mode which becomes unbounded, as 
energy density increases, is the mode $k = N/2 - 1$. A simple approximate formula, valid for large $N$ and 
$\mu = 1$ and derived approximating the Hill's matrix with a $3 \times 3$ matrix, gives for  the threshold energy 
density: \par
\begin{equation}
\epsilon_{t} = \frac{E_{t}}{N} = \frac{3.226}{N^{2}} + 0(N^{-4}).
\label{eq:et}
\end{equation}   
The problem of stability of the mode $N/2$ was also tackled in \cite{fla}, in connection with the study of tangent
bifurcation of band edge plane waves in nonlinear Hamiltonian lattices. In the limit of large $N$, the formula 
\begin{equation}
\epsilon_{t} = \frac{E_{t}}{N} = \frac{\pi^2}{3 N^{2}} \approx \frac{3.29}{N^{2}}
\label{eq:eps}
\end{equation}
is derived for the bifurcation energy density corresponding to the threshold energy density.  This result is slightly
different from the result (\ref{eq:et}) and  the small difference is probably due to the rough estimate of the 
eigenvalue spectrum of the Hill's matrix in \cite{bud}. 

The problem of stability of a OMS was reconsidered subsequently in \cite{poggi} with a detailed 
analysis of the mode $N/2$. This is the simpler case, because, as we have seen, the different components of the 
perturbation, in modal space, are all decoupled and can be reduced to the single Lam\'e equation:
\bigskip
\begin{equation}
\stackrel{..}x_{r} = -{\omega_{r}}^2 [1 + \frac{12 ~\mu ~ A^{2}~ cn^{2}(\Omega_{N/2}t;k)}{N}]x_{r}, ~~~r = 1,...,N-1.
\label{eq:lam}
\end{equation}
\bigskip
To obtain in very simple way this equation, we observe that from (\ref{eq:h}), (\ref{eq:h0}) and (\ref{eq:h1}) 
one has: 
\bigskip
\begin{equation}
\stackrel{\cdot}{p_{k}} = q_{k+1}+q_{k-1} -2q_{k} + \mu [(q_{k+1}- q_{k})^3 -(q_{k} - q_{k-1})^3].
\end{equation}
\bigskip 
If the coordinates $q_{k}$ are affected by some error, then the error on the $\stackrel{\cdot}{p_{k}}$,~ 
$\triangle \stackrel{\cdot}{p_{k}}$, will be:
\bigskip
\begin{equation}
\triangle \stackrel{\cdot}{p_{k}} = \triangle {q_{k+1}}+ \triangle {q_{k-1}} - 2\triangle {q_{k}} 
+ 3 \mu [(q_{k+1}- q_{k})^{2} (\triangle {q_{k+1}} - \triangle {q_{k}}) -(q_{k} - q_{k-1})^{2} 
(\triangle {q_{k}} -\triangle {q_{k-1}})].
\label{eq:tria}
\end{equation}
\bigskip
From (\ref{eq:k}) and (\ref{eq:q-q}) we have: \par
\begin{equation}
(q_{k+1} - q_{k})^{2} = (q_{k} - q_{k-1})^{2} = \frac{4}{N}~{Q_{\frac{N}{2}}}^2.
\label{eq:qkp1}
\end{equation}
\bigskip
\noindent From (\ref{eq:tria}) and (\ref{eq:qkp1}) we obtain:
\bigskip
\begin{equation}
\triangle \stackrel{\cdot}{p_{k}} = \triangle q_{k+1} + \triangle q_{k-1} - 2 \triangle q_{k} + 
\frac{12 \mu}{N} {Q^{2}}_{\frac{N}{2}} [\triangle q_{k+1} + \triangle q_{k-1} - 2 \triangle q_{k}].
\label{eq:trian}
\end{equation}
\noindent Since $\triangle \stackrel{\cdot}{q_{k}} = \triangle p_{k}$, ~(\ref{eq:trian}) reads:
\bigskip
\begin{equation}
\triangle \stackrel{\cdot \cdot}{q_{k}} = [1 + \frac{12 \mu}{N} {Q^{2}}_{\frac{N}{2}}] 
[\triangle q_{k+1} - 2 \triangle q_{k} + \triangle q_{k-1}].
\label{eq:triangle}
\end{equation}
\noindent Passing to modal variables $Q_{k}$, we finally have: 
\bigskip
\begin{equation}
\triangle \stackrel{\cdot \cdot}{Q_{k}} = - {\omega^{2}}_{k} [1 + \frac{12 \mu}{N} {Q^{2}}_{\frac{N}{2}}] 
 ~\triangle Q_{k},
\label{eq:triaq}
\end{equation}
\bigskip
which is equation (\ref{eq:lam}).

We want to remark, however, that eq. (\ref{eq:triaq}) (and then (\ref{eq:lam})), on which is based the study 
of the stability reported in \cite{poggi}, is the result of a linear analysis and consequently all its 
implications have only local value.

We outline now the stability analysis of the one-mode $N/2$, given in \cite{poggi} . Let be $k$ the modulus of 
the Jacobian elliptic functions. It is related to the product $ \beta = \epsilon \mu$ by the relation: \par
\begin{equation}
\beta = \frac{k^{2}}{1 - 2 k^{2}} (1 + \frac{k^{2}}{1 - 2 k^{2}}).
\label{eq:betaf}
\end{equation}
\bigskip
Expressing $\Omega_{N/2}$ and $\mu A^{2}/N$, as functions of $k$, and after rescaling time at each fixed $k$, 
each of the Lam\'e equations (\ref{eq:lam}) can be put in the standard form: \par
\begin{equation}
y" + [~\alpha - \nu (\nu +1) ~k^{2} ~sn^{2} (u,k)~]~y = 0, 
\label{eq:y}
\end{equation}
where the prime superscript denotes differentiation with respect to the new time $u = \Omega_{N/2} t$,  $y$ 
stands for the generic variable $x_{r}$ and the parameters $\alpha$ and $\nu$ are defined by the 
relations: \par
\[  \begin{array}{ll} 
\alpha = \frac{1}{4} ( 1 + 4 k^{2}){\omega^{2}}_{r}, ~~~~~  \nu (\nu + 1) = \frac{3}{2}{\omega^{2}}.
     \end{array}
     \]
Putting: 
\begin{equation}
\rho = \sin^{2}(\pi ~r / N),
\label{eq:rho}
\end{equation}
\noindent the two last relations become: \par
\[  \begin{array}{ll} 
\alpha = \rho ( 1 + 4 k^{2}), ~~~~~  \nu (\nu + 1) = 6 \rho.
     \end{array}
      \]     
A given pair $(\rho, ~k)$ describes different mode numbers $r$ in systems with different $N$, all having 
the same $r/N$ value, with the same value of the product $\beta = \epsilon \mu$ for the unperturbed solution 
$n = N/2$. 
To solve the stability problem for generic values of $\nu$, the equation (\ref{eq:y}) is reduced to a Mathieu 
equation, by approximating the $sn^{2}$  by its first-order Fourier expansion. After a further 
time rescaling $\tau = \pi ~t' /2~K$, where $K$ is the complete elliptic integral of the first kind, with 
modulus $k$, one obtains the canonical form of Mathieu equation: \par
\begin{equation}
\frac{d^{2} y}{d \tau^{2}} + [ a - 2~q \cos {(2~\tau) } ]~y = 0,
\label{eq:frac}
\end{equation}
\noindent where: \par
\begin{equation}
a = \rho (\frac{2 K}{\pi})^{2} ( -5 + 4 k^{2} + 6 \frac{E}{K}),
\label{eq:arho}
\end{equation}
\par
\begin{equation} 
q = - \frac{12 ~\rho}{\sinh{(\pi K'/K)}},
\label{eq:qfrac}
\end{equation} 
\begin{equation}
K'(k) = K \sqrt{1 - k^{2}},
\label{eq:k'k}
\end{equation}
\par
\noindent and $E$ is the complete elliptic integral of the second kind, with modulus $k$. Both parameters $a$ and 
$q$ depend on $k^2$ and $\rho$. For each fixed $\rho$, which corresponds to fix a ratio $r/N$, changing $k^2$, 
namely $\beta$, one can trace a curve in the $(q, a)$ parameter plane. A stability transition happens 
if this curve intersects characteristic curves separating stable from unstable regions for the Mathieu 
equation. \par 
The main results of the stability analysis reported in \cite{poggi} are: \par
a) for each mode having $\rho > 1/3$, there is a threshold value of $\beta$ for instability. 
Above this threshold the nonlinear mode $n = N/2$ presents an instability causing growth of the mode corresponding 
to $\rho$, through parametric resonance; \par
b) on the contrary, modes with $\rho < 1/3$ ~(i.e. $r/N < 0.196$) ~are always stable in the linear approximation, 
for any energy density of mode $n = N/2$, so that, perturbations of this mode involving only modes with 
$\rho < 1/3$ never lead to instability. These modes, as well as modes with $\rho > 1/3$, when the product 
$\epsilon \mu$ is less than the critical value, can grow only if they are triggered by the interaction with other 
modes that are unstable; \par
c) for $N \geq 4$ there are always modes with $\rho > 1/3$ so that the mode $N/2$ can never be stable for all energy 
densities. Since the threshold value of $\beta$ is a decreasing function  of $\rho$, the first modes to go  
unstable, when $\epsilon \mu$ is increased from zero, are the modes ~$r = N/2 -1$~ and ~$r = N/2 +1$, ~which have 
$\rho = \cos^{2}(\pi / N)$. Therefore, for each (even) number $N$ of particles, there is a non-zero value of 
$\beta$, function of $N$, below which the nonlinear mode $N/2$ is stable. Since the critical value tends to 
zero for $\rho \rightarrow 1$, the critical threshold for the stability approaches zero as the number $N$ of 
particles is increased without limit; \par
d) using power series expansion of the various previous formulas, one obtains, for the critical value of threshold, 
the formula: \par
\begin{equation}
\beta_{t} = \frac{\pi^{2}}{3 N^{2}} + 0( N^{-4} ),
\label{eq:beta}
\end{equation}

\noindent which confirms the $N^{-2}$ dependence found in \cite{bud} and in \cite{fla} and the numerical values 
$\pi^{2}/3$ of the coefficient of $1/N^{2}$ found in \cite{fla}.  

\bigskip
\section{Numerical results for the case $N/2$}

In this section, we present the results of our numerical analysis of the stability of the one-mode 
solution corresponding to $n = N/2$, as a function of the product $\epsilon \mu$ and of the number $N$ 
of particles, based on the numerical integration of the  full nonlinear FPU model directly in the variables 
$q_{k}$, $p_{k}$. More precisely, we integrate the equations of motion in the variables $q_{k}, p_{k}$ by means of 
a bilinear symplectic algorithm of the third order, which has been adapted from an algorithm employed previously 
by Casetti \cite{cas}. Initial conditions for the variables $q_{k}, p_{k}$ are obtained in the following way.  
We excite the OMS, at $t = 0$, always putting $Q_{N/2} \neq 0$ and $P_{N/2} = 0$. In all the 
numerical experiments  we fix $\mu = 0.1$ and change the value of the energy density 
$\epsilon = E_{N/2}/N$ where  
\begin{equation}
E_{N/2} = \frac{1}{2} (P_{N/2}^{2} + \omega_{N/2}^{2}~ Q_{N/2}^{2} + \mu~ \frac{\omega_{N/2}^{4}~Q_{N/2}^{4}}{2~N})
\label{eq:enme}
\end{equation}
\noindent is the energy of the nonlinear one-mode $N/2$. If we fix the initial value of $E_{N/2}$ (or 
equivalently $\epsilon$), the initial value of $Q_{N/2}$ is obtained from (\ref{eq:enme}) with $P_{N/2} = 0$. 
Finally, from inverse transformations of (\ref{eq:qk}) and (\ref{eq:pk}), the values of $q_{k}(0)$ and  $p_{k}(0)$ 
are obtained.

The normal coordinates $Q_{N/2}$ and $P_{N/2}$ of the nonlinear one-mode and the normal coordinates $Q_k$ and $P_k$ 
of the other  normal modes are calculated at fixed time intervals, multiple of the integration step. Then, the 
study of the stability of the OMS is made through the analysis of both the time evolution of $Q_{N/2}$ and $P_{N/2}$  
and the evolution of the other modes $Q_{k}$, $P_{k}$ which are  generated through computational errors. 

As concerns the numerical integration, we use values of integration time steps $\triangle t$ ranging from 
$0.01$ to $0.005$. For example, the first value is approximately $1/300$ of the smallest period of oscillation 
of the atoms in the harmonic case  and  allows us to obtain a control of the total energy $E$ of the 
lattice, which ensures a relative error $\triangle E/ E< 10^{-6}$. 

To illustrate the various steps of our method of numerical analysis, let us consider the case $N = 32$ and thus the 
OMS $N/2 = 16$. In this case $C_{nnnn} = 2$ and we have the formula (\ref{eq:enme}) for the energy of the 
nonlinear mode. We take a value of the energy density $\epsilon$ and integrate the equations of motion for the 
variables $q_{k}$ and $p_{k}$. The integration time is fixed in such a way to observe the instability of nonlinear 
mode, if the value of $\beta = \epsilon \mu$ is greater than the theoretical value (\ref{eq:beta}) of the threshold 
energy density. Typical values of this time are of order $10^{6} ~\triangle t$ where $\triangle t$ is the integration 
step. 

\begin{figure}[bhtp]
\centerline{\includegraphics[angle=-90,width=.4\textwidth]{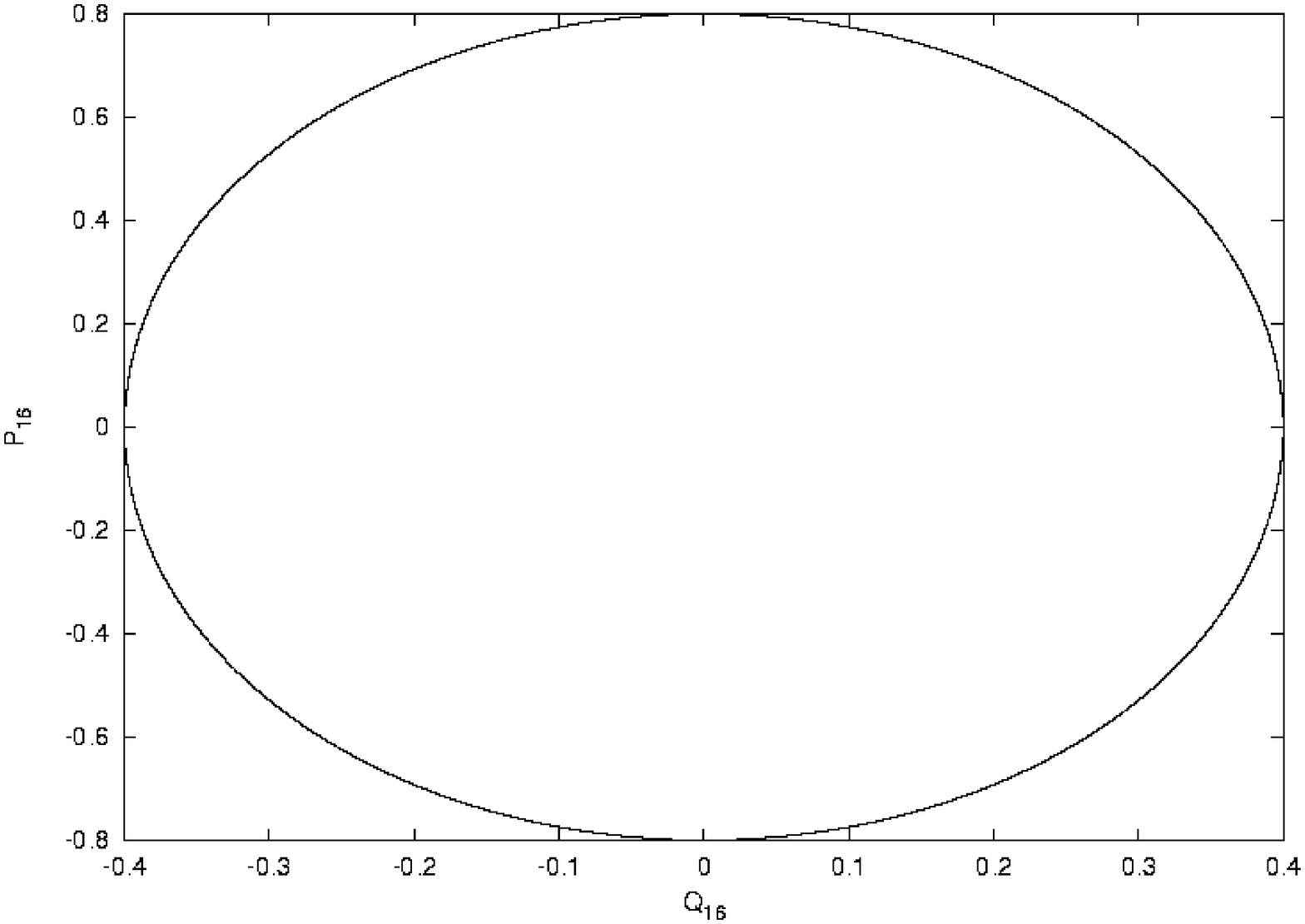}}
\caption{$P_{16}$ vs $Q_{16}$ for $\epsilon\mu = 0.001$}
\label{fi:qp01}
\centerline{\includegraphics[angle=-90,width=.4\textwidth]{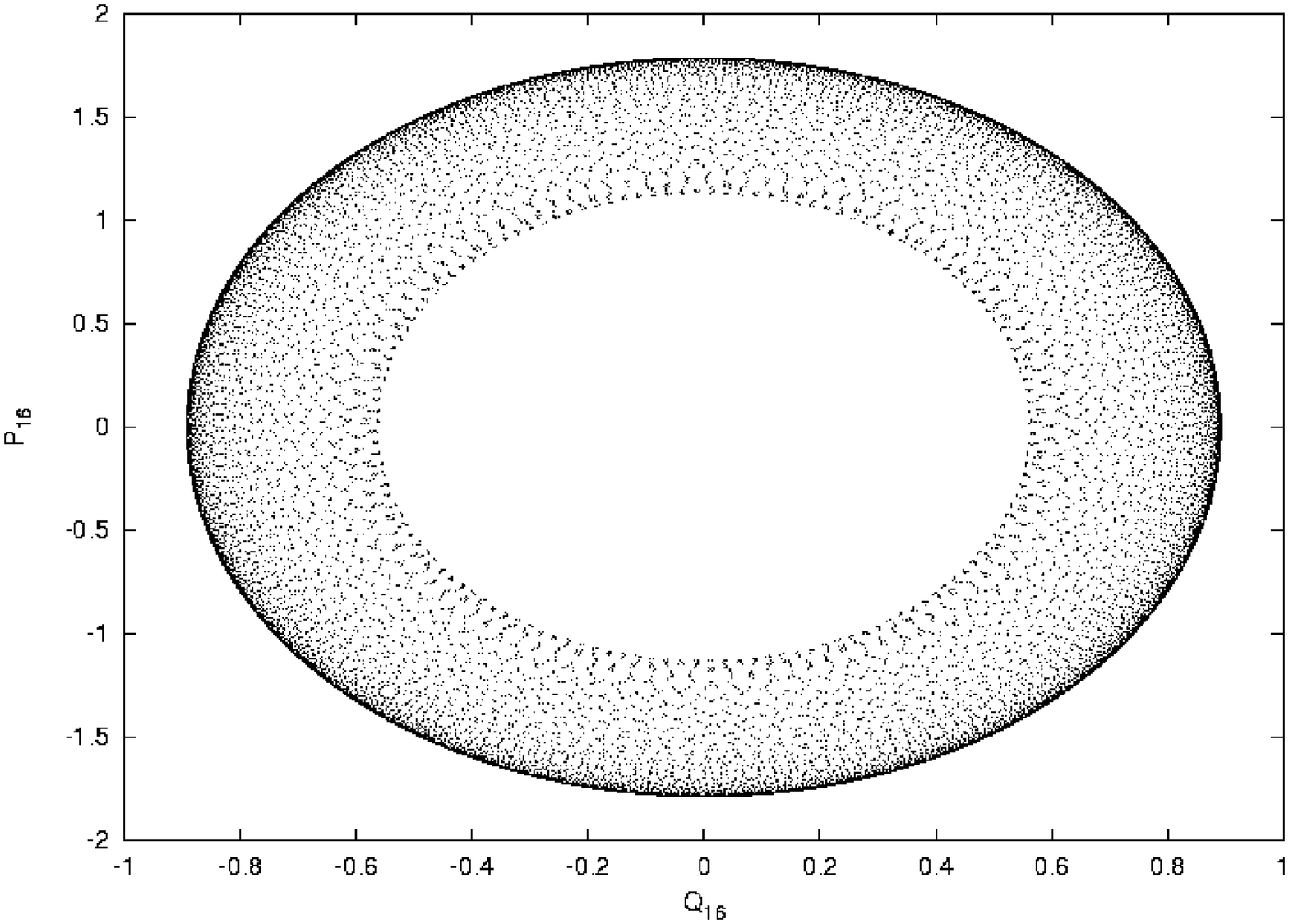}}
\caption{$P_{16}$ vs $Q_{16}$ for $\epsilon\mu = 0.005$}
\label{fi:qp5}
\centerline{\includegraphics[angle=-90,width=.4\textwidth]{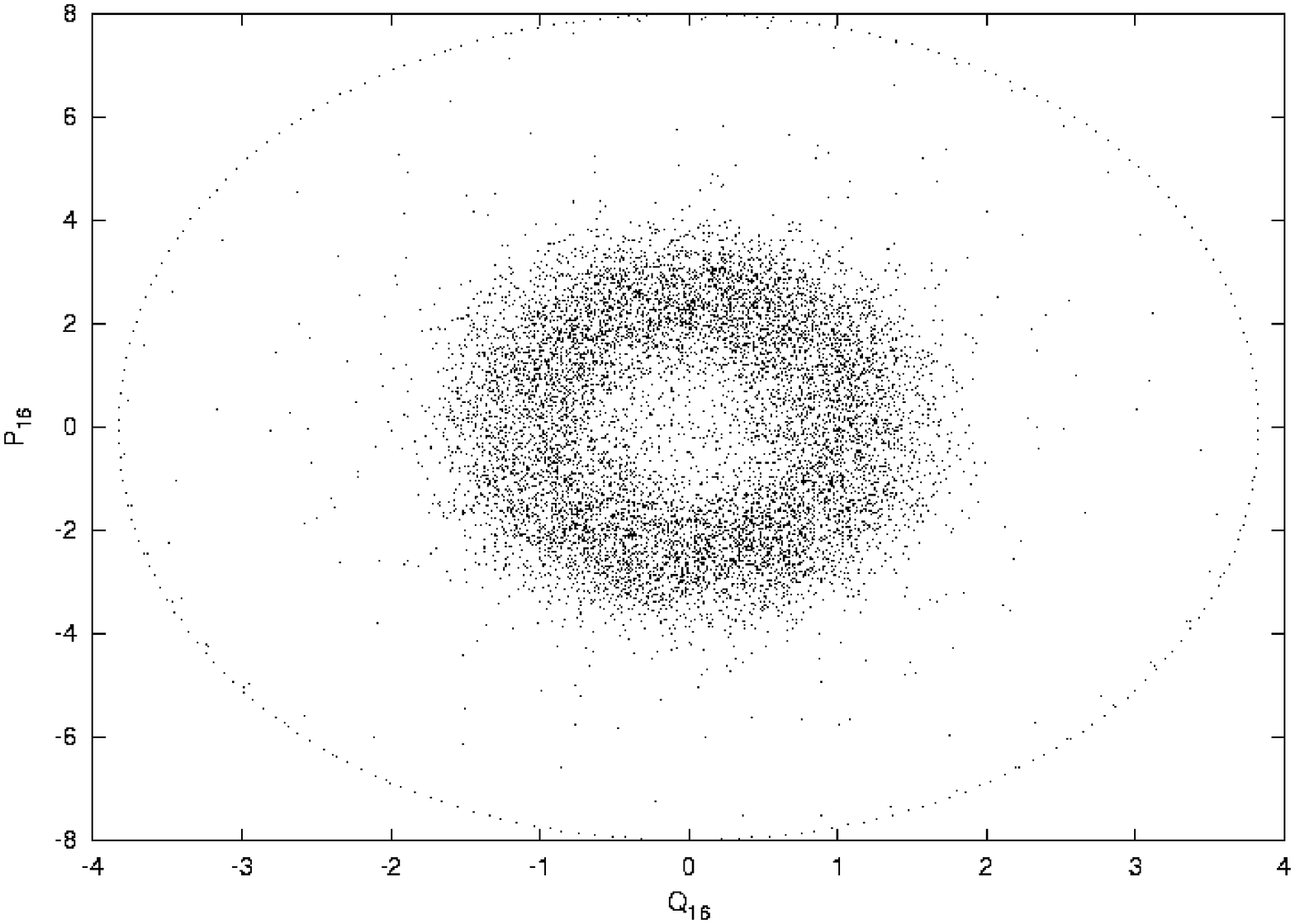}}
\caption{$P_{16}$ vs $Q_{16}$ for $\epsilon\mu = 0.1$}
\label{fi:qp1}
\end{figure}

\begin{figure}[htbp]
\centerline{\includegraphics[angle=-90,width=.5\textwidth]{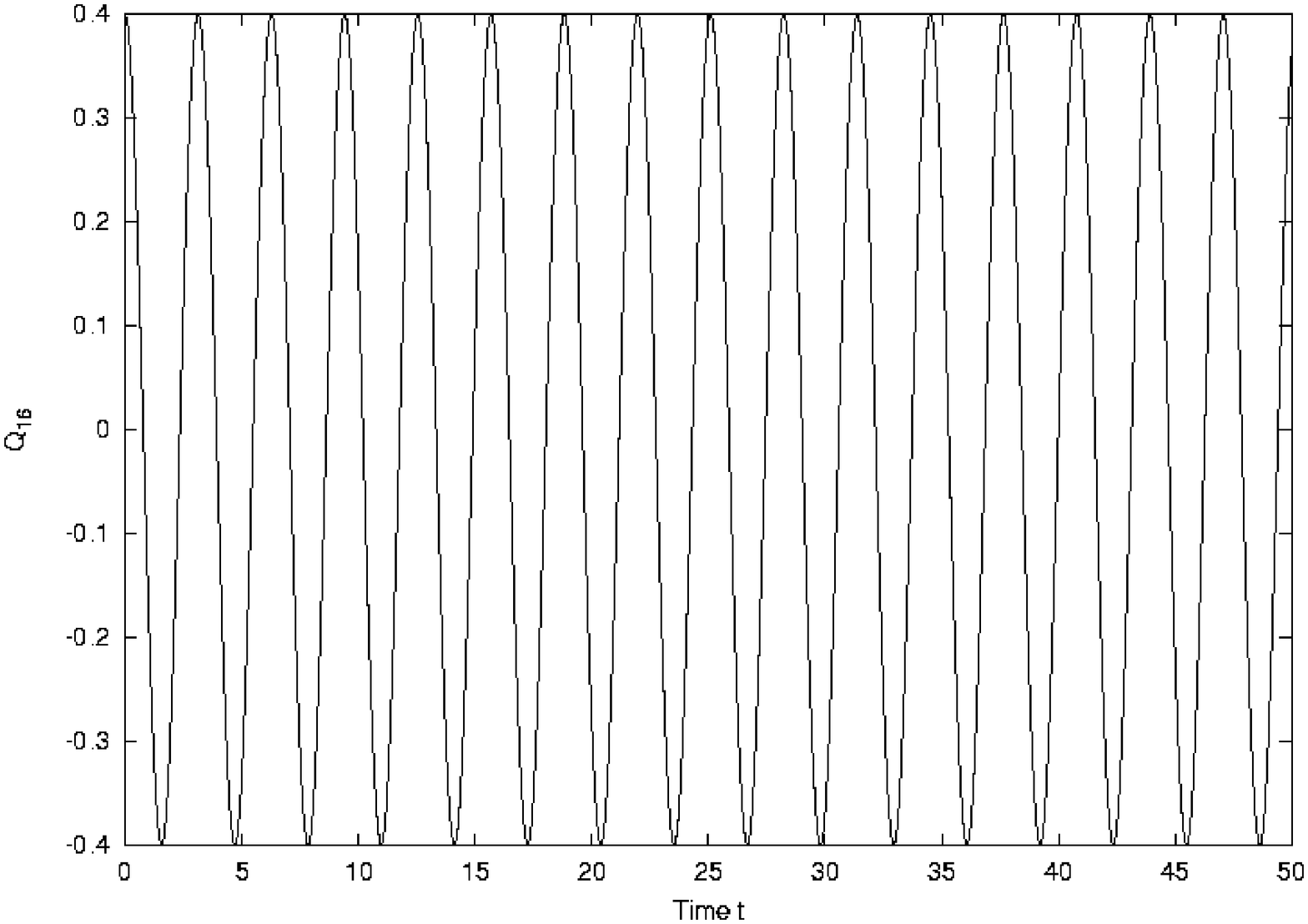}}
\caption{$Q_{16}$ vs $t$ for $\epsilon\mu = 0.001$}
\label{fi:q01}
\centerline{\includegraphics[angle=-90,width=.5\textwidth]{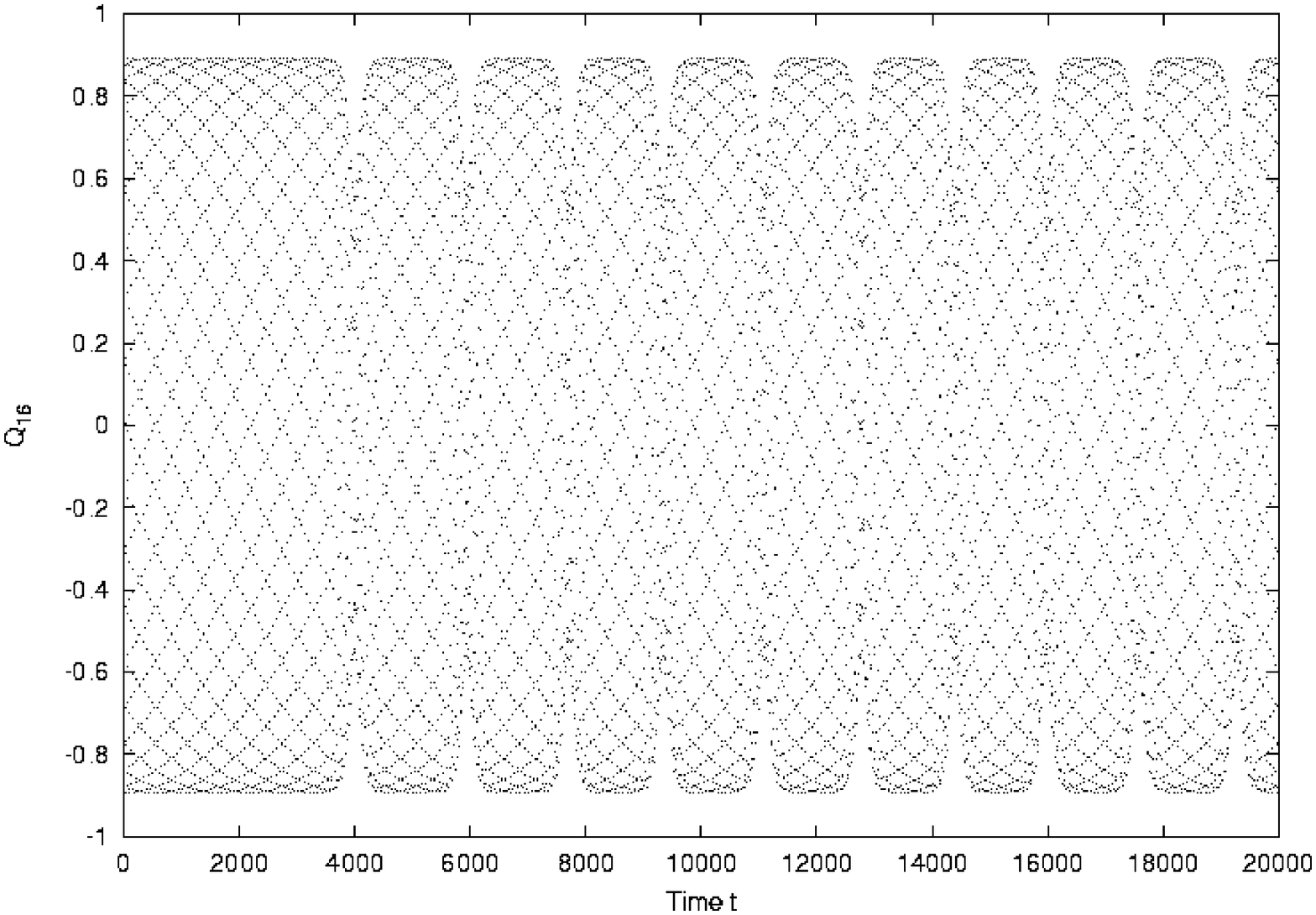}}
\caption{$Q_{16}$ vs $t$ for $\epsilon\mu = 0.005$}
\label{fi:q5}
\centerline{\includegraphics[angle=-90,width=.5\textwidth]{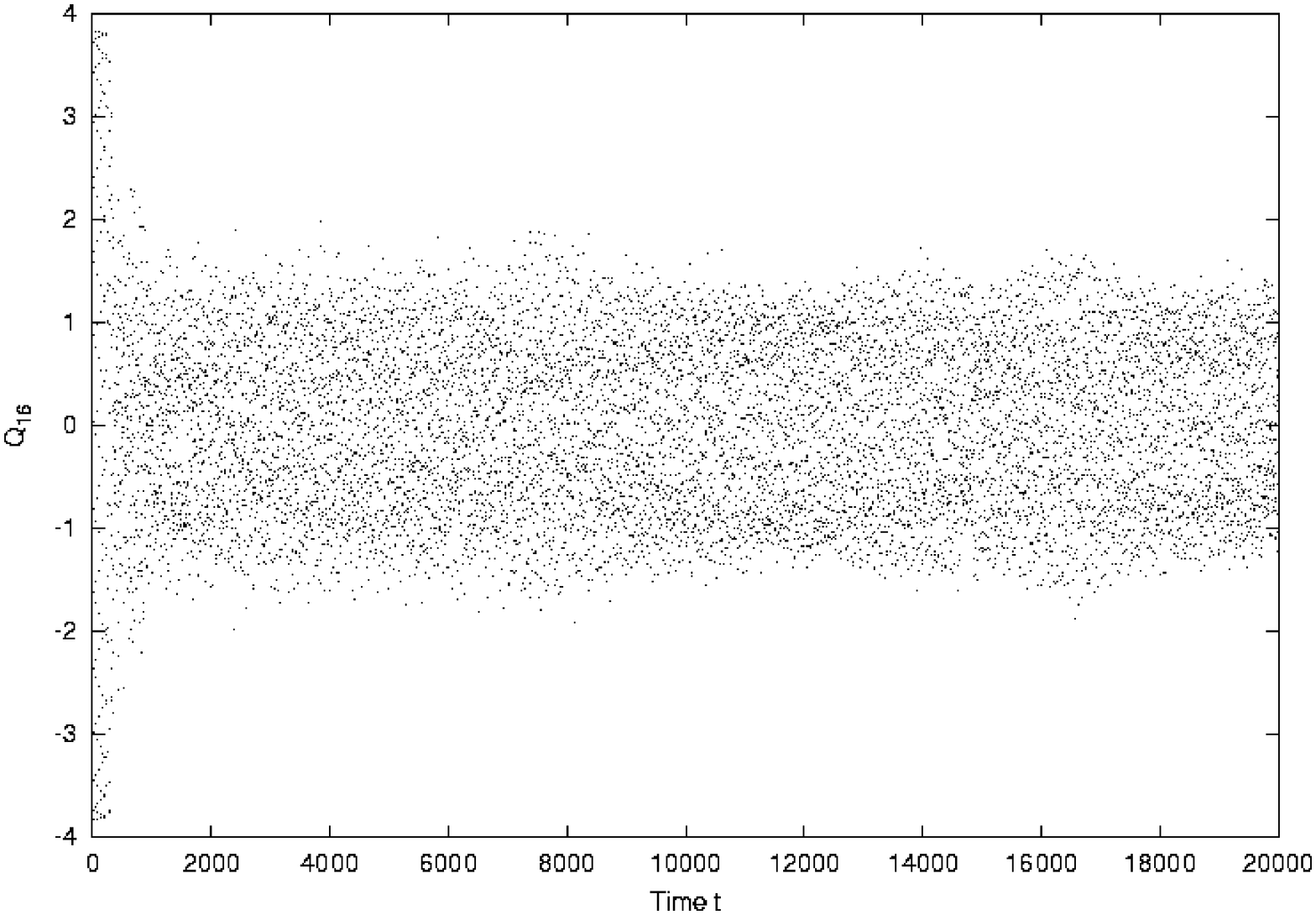}}
\caption{$Q_{16}$ vs $t$ for $\epsilon\mu = 0.1$}
\label{fi:q1}
\end{figure}

For $N = 32$, we have $\beta_{t} = \pi^{2}/3 N^2 = 0.00321$. We consider three values of $\beta$: the first one, 
$\beta = 0.001$, smaller than $\beta_{t}$, the second one, $\beta = 0.005$, larger and the third, $\beta = 0.1$,  
much larger than $\beta_{t}$. In Figs. \ref{fi:qp01} - \ref{fi:qp1},the behaviours  of the nonlinear mode in the plane 
$Q_{16}, P_{16}$ are shown for these three values of $\beta$, while in Figs. \ref{fi:q01} - \ref{fi:q1}
we  show $Q_{16}(t)$, as a function of time  for the same values of $\beta$.
 
\eject

\begin{figure}[htbp]
\centerline{\includegraphics[angle=-90,width=.6\textwidth]{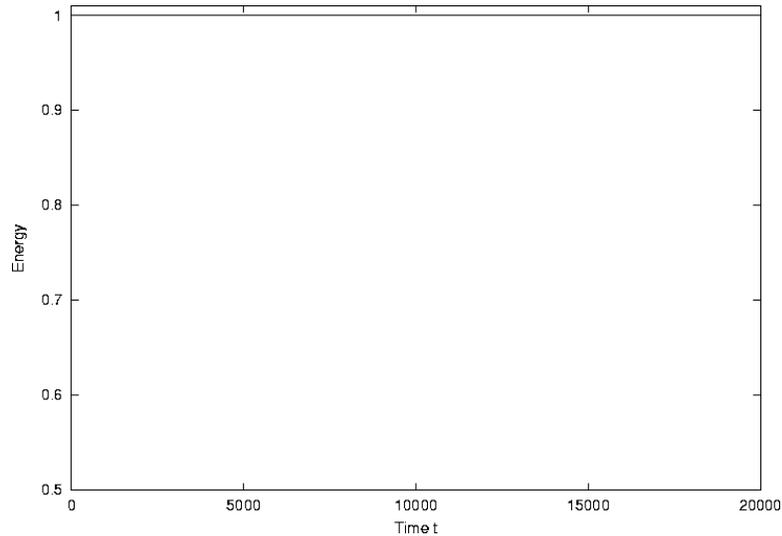}}
\caption{Energy of the mode $16$ vs $t$ for $\epsilon\mu = 0.001$}
\label{fi:qe001}
\end{figure}

\begin{figure}
\centerline{\includegraphics[angle=-90,width=.6\textwidth]{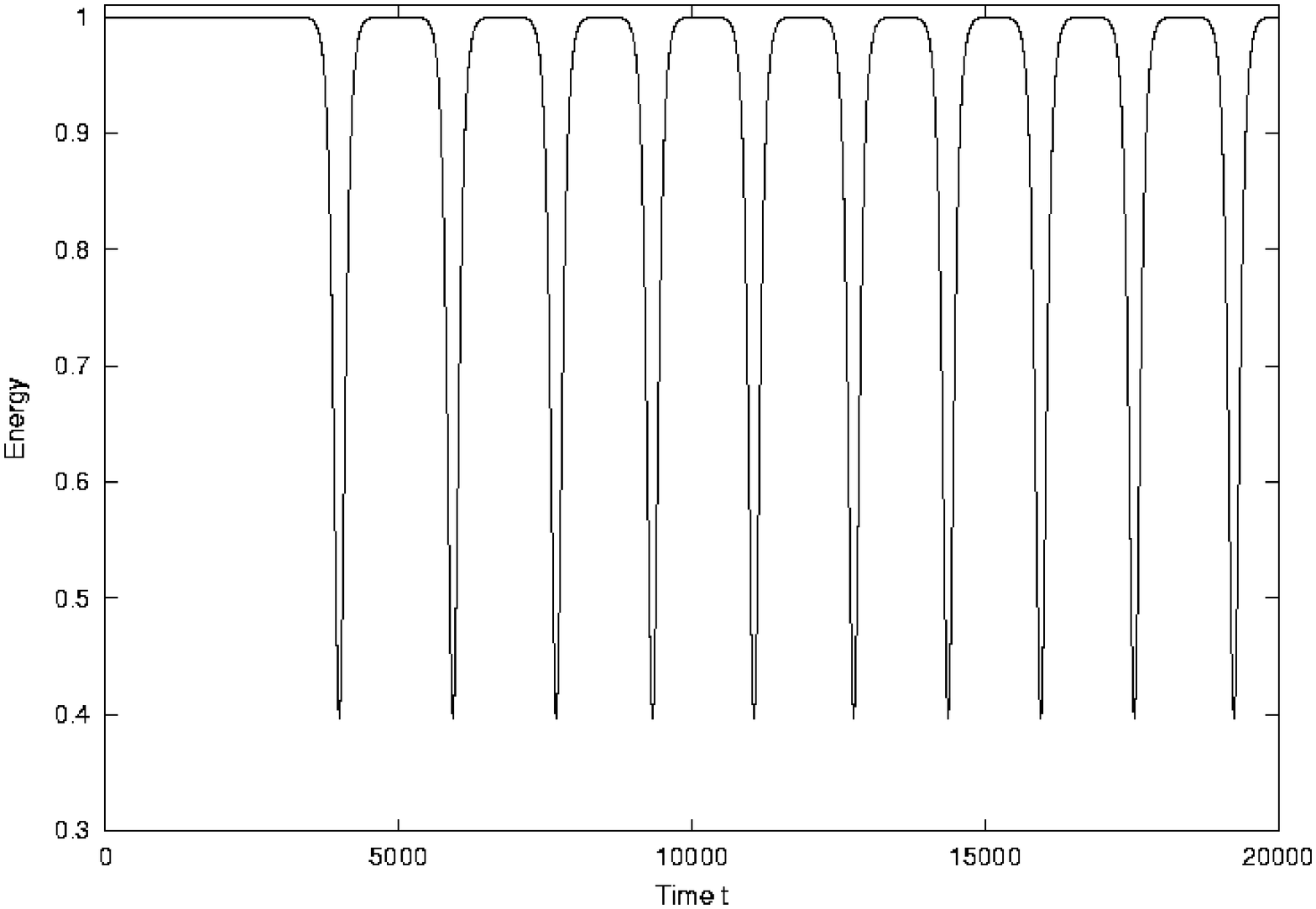}}
\caption{Energy of the mode $16$ vs $t$ for $\epsilon\mu = 0.005$}
\label{fi:ener1}
\centerline{\includegraphics[angle=-90,width=.6\textwidth]{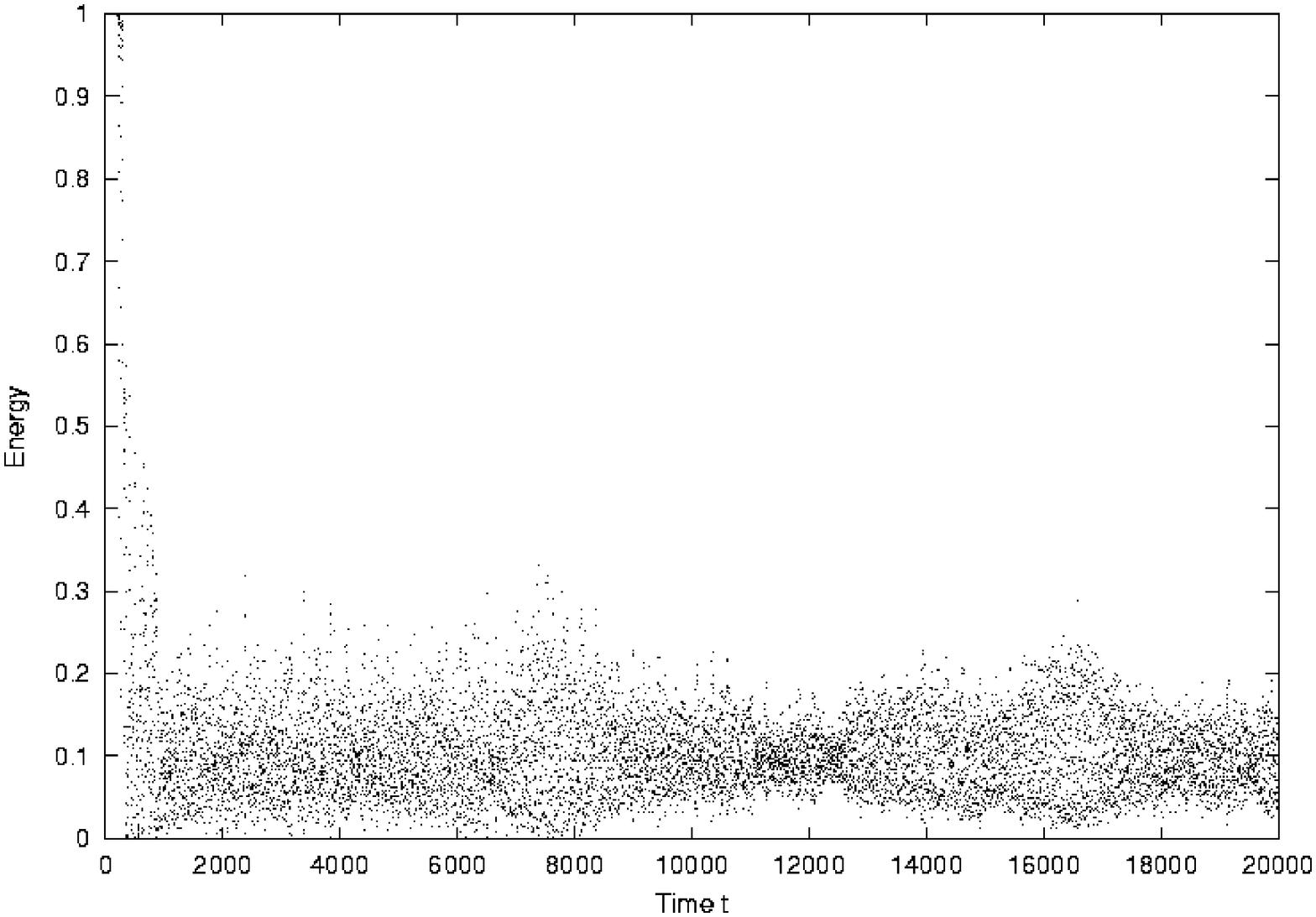}}
\caption{Energy of the mode $16$ vs $t$ for $\epsilon\mu = 0.1$ }
\label{fi:nm4}
\end{figure}

From the inspection of previous figures, we can observe that, for $\beta = \epsilon\mu$ very small, well below 
the threshold energy density, the nonlinear OMS $N/2$ is stable and $Q_{16}$ is a periodic function with 
the same amplitude and the same period of the analytical solution. For $\epsilon \mu =0.005$, above 
the threshold, the situation is very different. The period of oscillation is equal to the period of the analytical 
solution, and for very long times, in the plane $Q_{16}, P_{16}$, the representative point moves on a one-nonlinear 
dimensional 
closed curve, as for values of $\epsilon \mu$ very small; but now, periodically and for short times, the amplitude of 
oscillation varies, due to a decrease of modal energy, and the representative point 
of the system moves on an open curve  which tends periodically to shrink. For $\epsilon \mu =0.1$, well above the 
threshold energy density, we observe a chaotic behaviour. This behaviour is also evident from the inspection of the 
figures \ref{fi:qe001} - \ref{fi:nm4} in which the modal energy is shown, as a function of time, for the three 
values of $\beta$. In these figures, and in all the next figures which show the behaviour of energy vs time, 
the energy is always normalized to the initial excitation energy.

\bigskip 
\eject
\bigskip

\begin{figure}[thbp]
\centerline{\includegraphics[angle=-90,width=.6\textwidth]{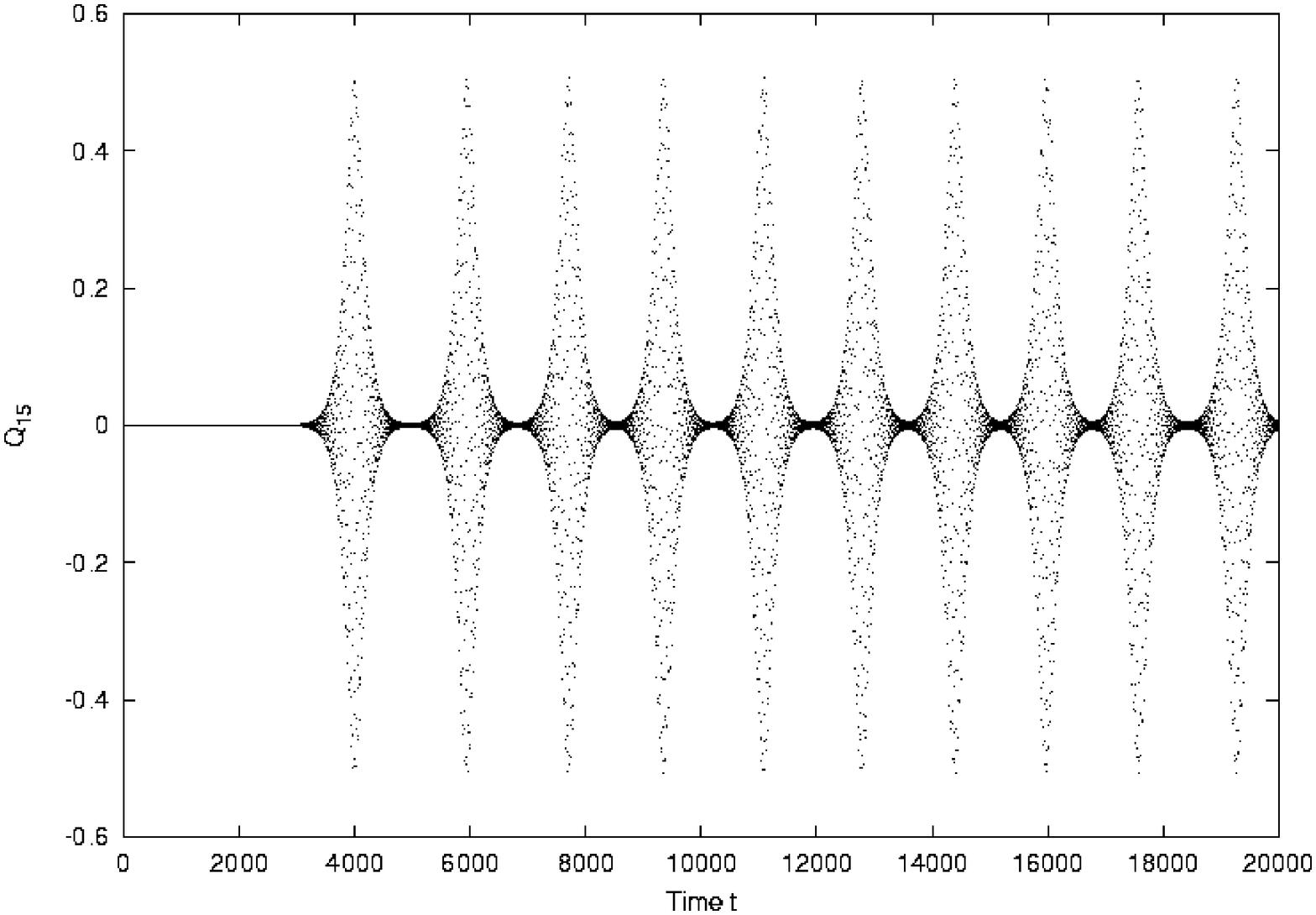}}
\caption{$Q_{15}$ vs $t$ for $\epsilon\mu = 0.005$}
\label{fi:q15}
\centerline{\includegraphics[angle=-90,width=.6\textwidth]{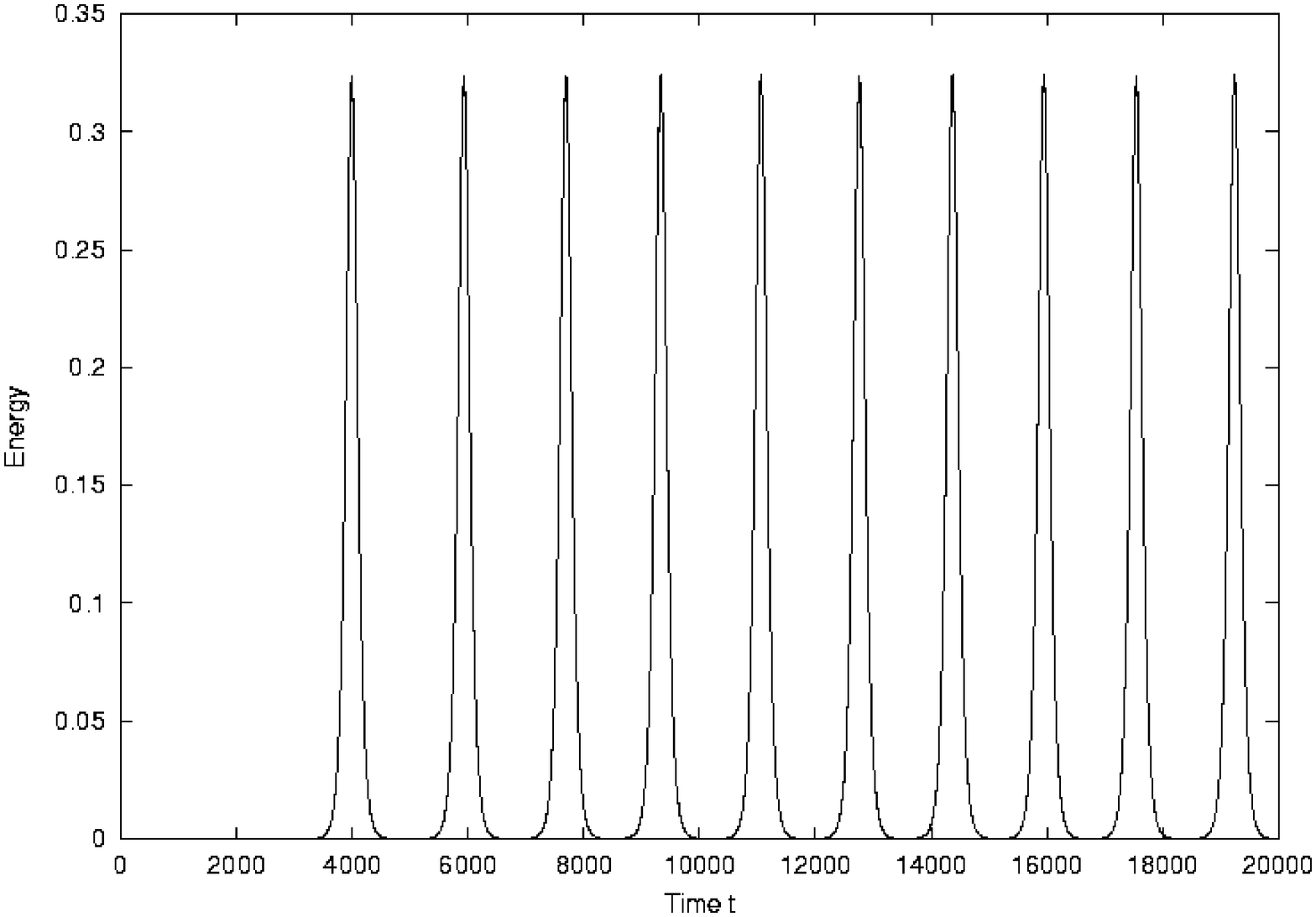}}
\caption{Energy of the mode $15$ vs $t$ for $\epsilon\mu = 0.005$}
\label{fi:eq15}
\end{figure}

As can be seen from the last figures, it is evident that the mode $16$ exchanges energy with some other mode.
According to the theory developed in \cite{bud} and in \cite{poggi}, for $\beta = 0.005$, we are above the 
threshold for the excitation of the adjacent mode $15$, which is the first to be excited, and the excitation of 
this mode, due to nonlinear coupling between the modes, should trigger other linear modes.
In the figures \ref{fi:q15} and \ref{fi:eq15} the behaviour of the the adjacent mode $15$ is shown. \par
\begin{figure}[htbp]
\centerline{\includegraphics[angle=-90,width=.6\textwidth]{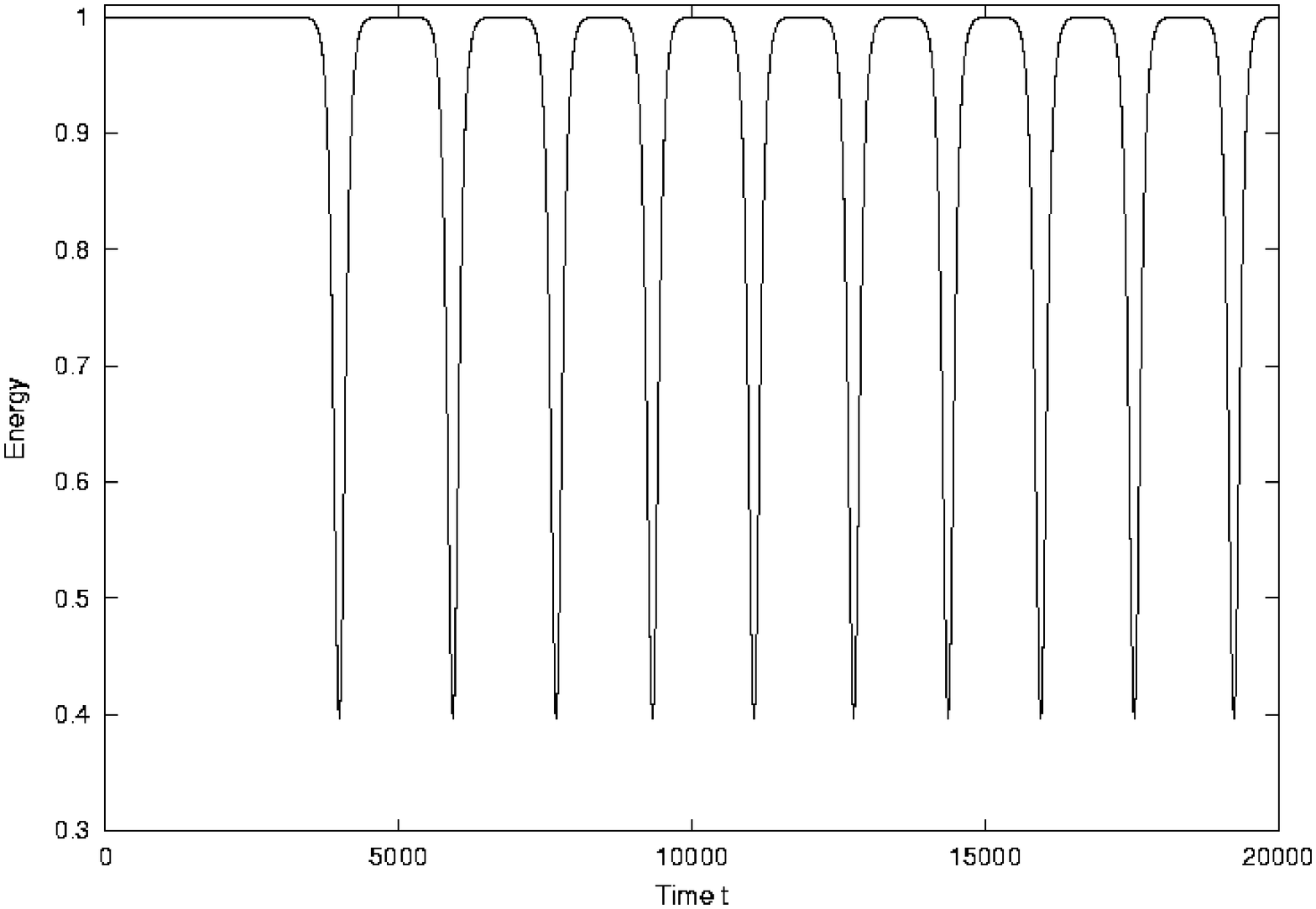}}
\caption{Energy of the mode $16$ vs $t$ for $\epsilon\mu = 0.005$}
\label{fig:ener1}
\centerline{\includegraphics[angle=-90,width=.6\textwidth]{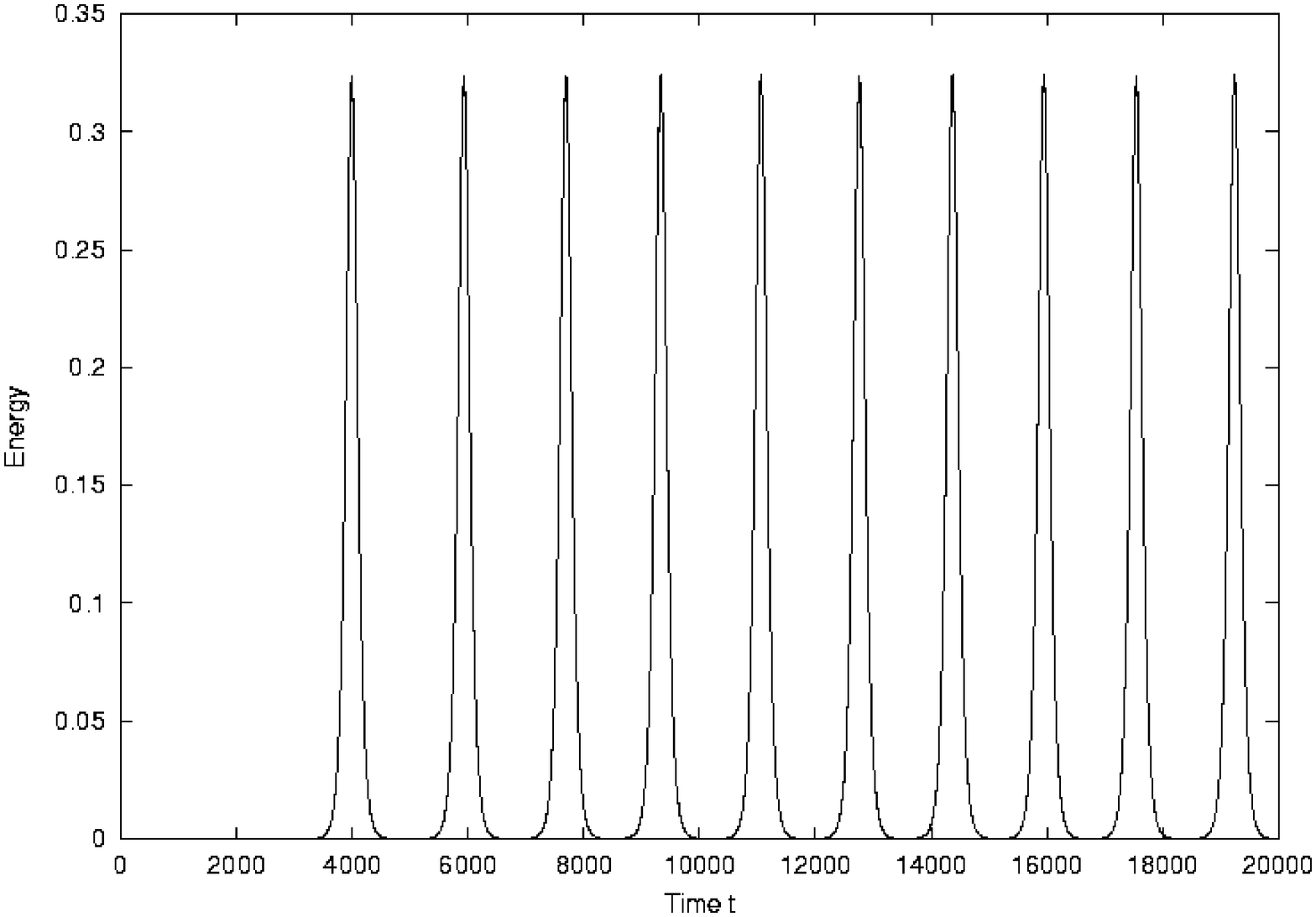}}
\caption{Energy of the mode $15$ vs $t$ for $\epsilon\mu = 0.005$}
\label{fig:eq15}
\centerline{\includegraphics[angle=-90,width=.6\textwidth]{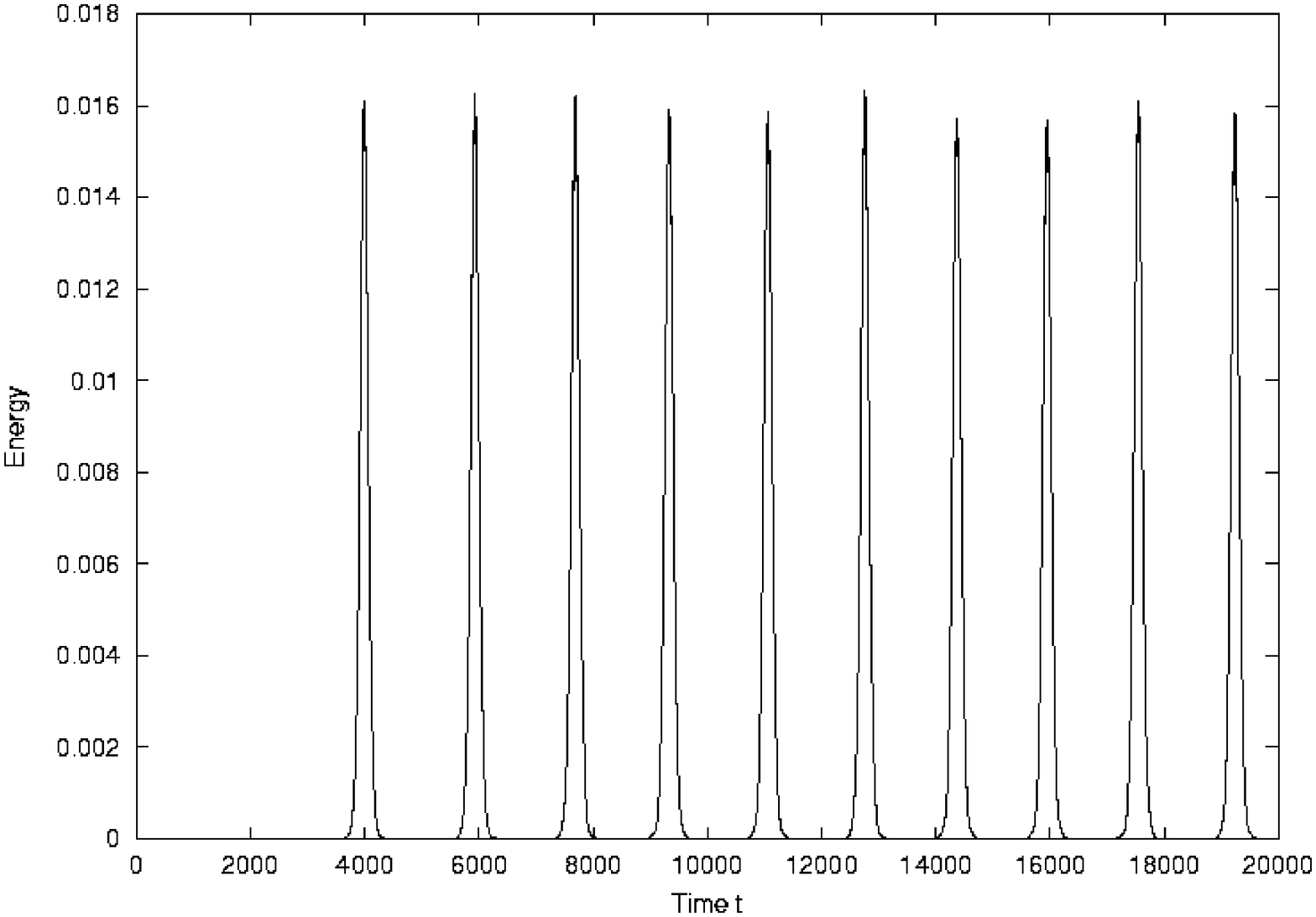}}
\caption{Energy of the mode $14$ vs $t$ for $\epsilon\mu = 0.005$}
\label{fig:ener2}
\end{figure}

\begin{figure}[htbp]
\centerline{\includegraphics[angle=-90,width=.6\textwidth]{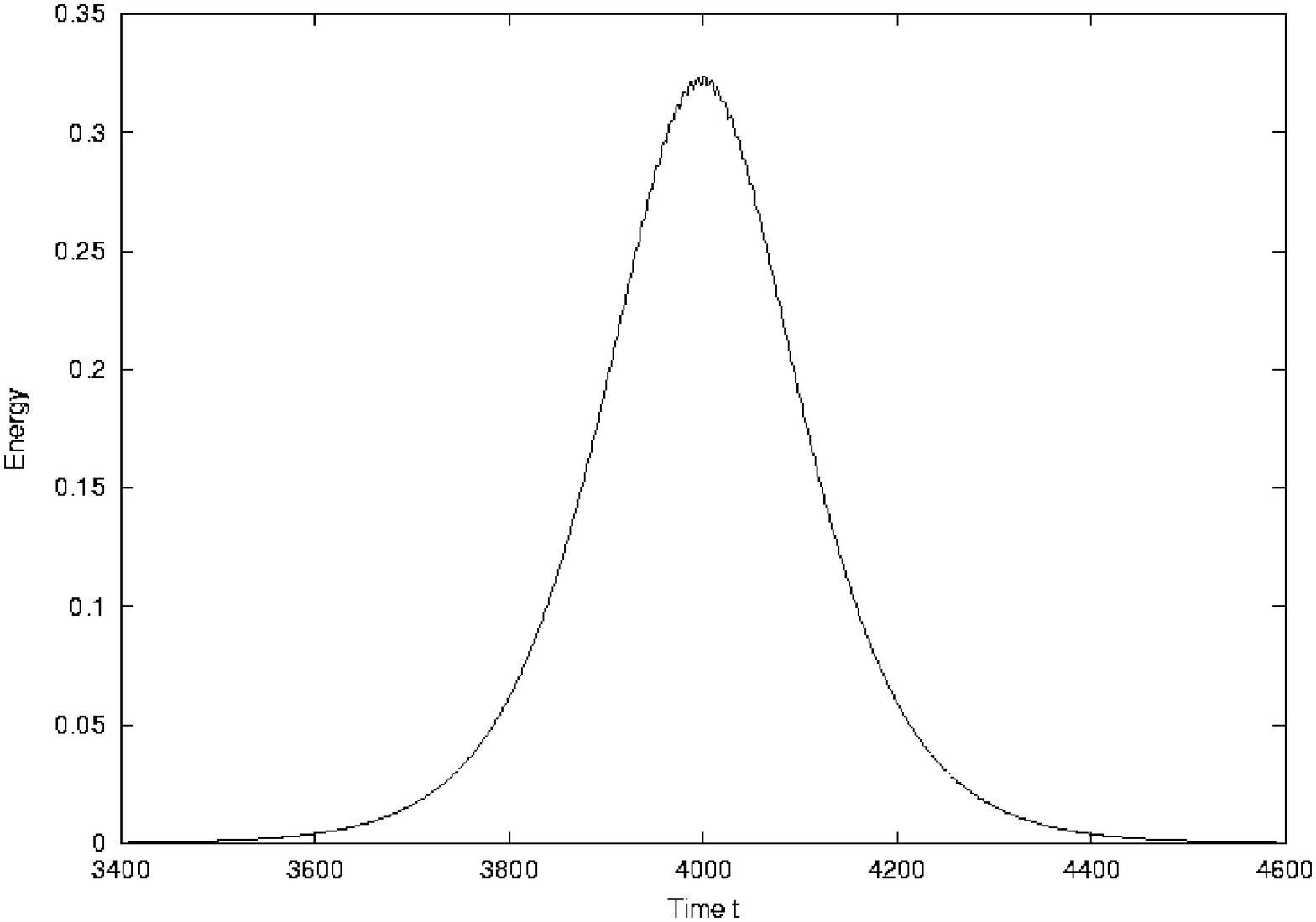}}
\caption{Behaviour of the energy of the mode $15$ vs $t$  for $\epsilon\mu = 0.005$ }
\label{fig:qener15}
\centerline{\includegraphics[angle=-90,width=.6\textwidth]{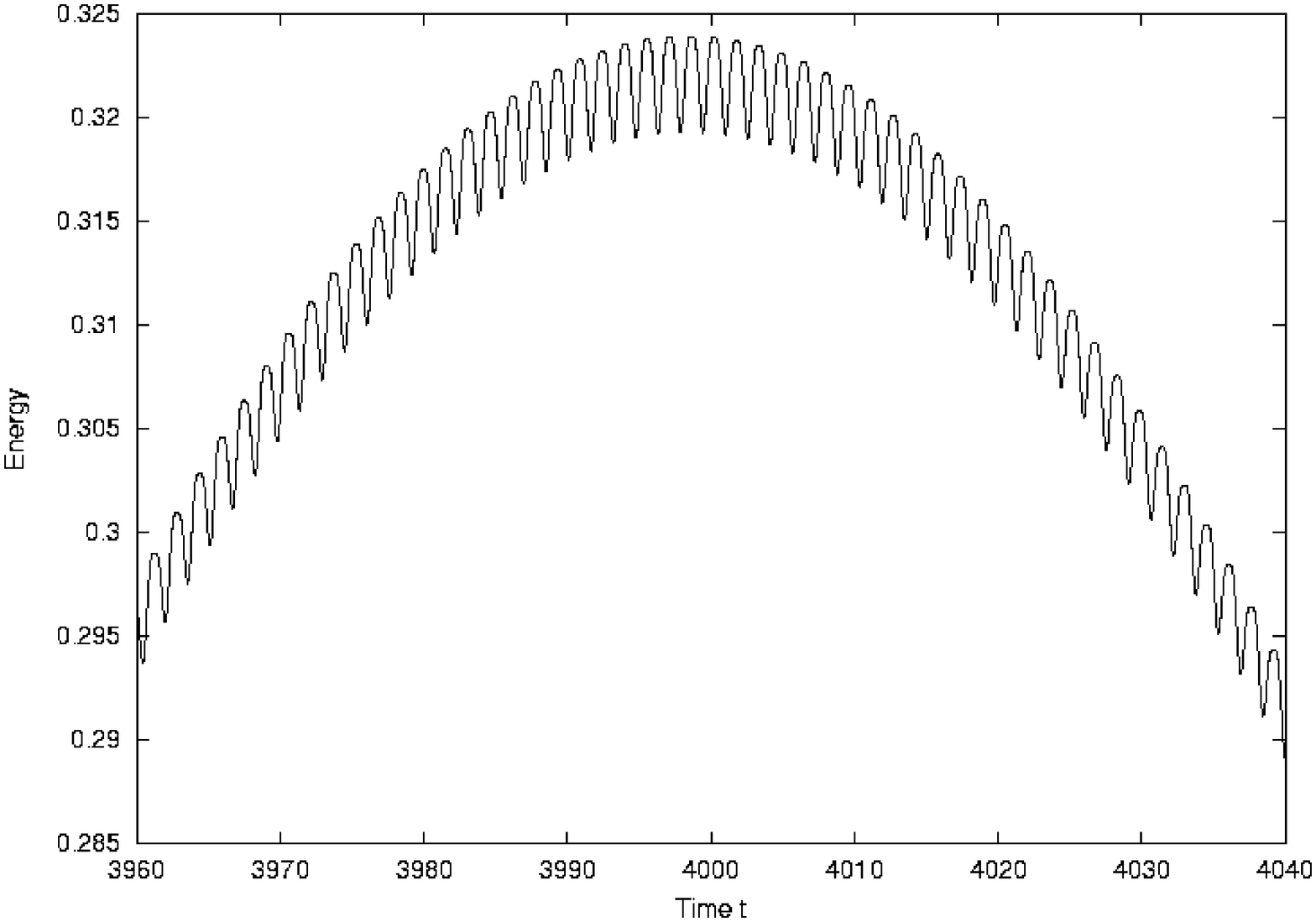}}
\caption{Behaviour of the energy of the mode $15$ vs $t$ for $\epsilon\mu = 0.005$, near the maximum}
\label{fig:qener15m}
\end{figure}

\eject 
The contemporary excitation of the other modes is shown in Figs. \ref{fig:ener1} - \ref{fig:ener2}, where the 
energies of the modes $n = N/2$, $n = 15$, $n=14$ are shown, as  functions of time.    
We remark that the energy of the mode $16$ is calculated with the formula (\ref{eq:en}),  while, for the other 
modes, the usual formula (\ref{eq:el}) is utilized. Of course, formula (\ref{eq:el}) is only indicative for large 
excitation energy, when the variables $Q_{k}$ and $P_{k}$ lose their meaning of modal variables. 
\vfill\eject
The first energy pulse, for the mode $15$, appears as in  Fig. \ref{fig:qener15}.  Near the maximum, 
the value of energy oscillates as in Fig. \ref{fig:qener15m} with a period $T \approx \pi/2$, equal to half 
the period of the oscillations of  $Q_{16}$. \par 
\vfill\eject

The initial time interval, necessary to excite, through computational errors, the other modes, depends obviously, 
on the precision of numerical computations. All the previous numerical calculations have been performed in double 
precision. We have observed that, working in simple precision, the exchange of energy of the mode $N/2$ with the 
other modes occurs much before than in double precision. However, since the mechanism is primed, it repeats with 
the same properties either in double or in simple precision.

\begin{figure}[htbp]
\centerline{\includegraphics[angle=-90,width=.7\textwidth]{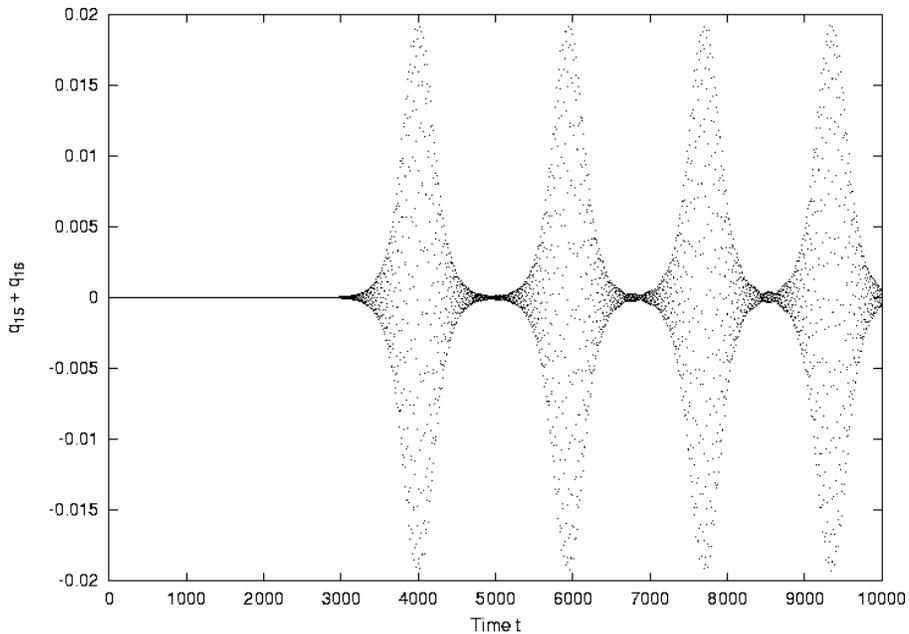}}
\caption{$q_{15} + q_{16}$ vs $t$ for $\epsilon\mu= 0.005$ when the mode $16$ is initially excited.}
\label{fig:coor1}
\end{figure}

The instability of the OMS can also be seen from another point of view, by considering the hamiltonian variables 
$q_{k}$. We recall that, if the OMS ~$Q_{N/2}$ were stable then, from (\ref{eq:q-q}), the sum $q_{k} + q_{k+1}$ 
would be always zero. In Fig. \ref{fig:coor1}, as an example, we report the sum $q_{15} +q_{16}$ as a function of 
time. As can be seen from the figure, the instability (sum of the two coordinates different from zero) appears 
when the mode $16$ starts to exchange energy with the other modes.
\bigskip
\eject

\begin{figure}[htbp]
\centerline{\includegraphics[angle=-90,width=.6\textwidth]{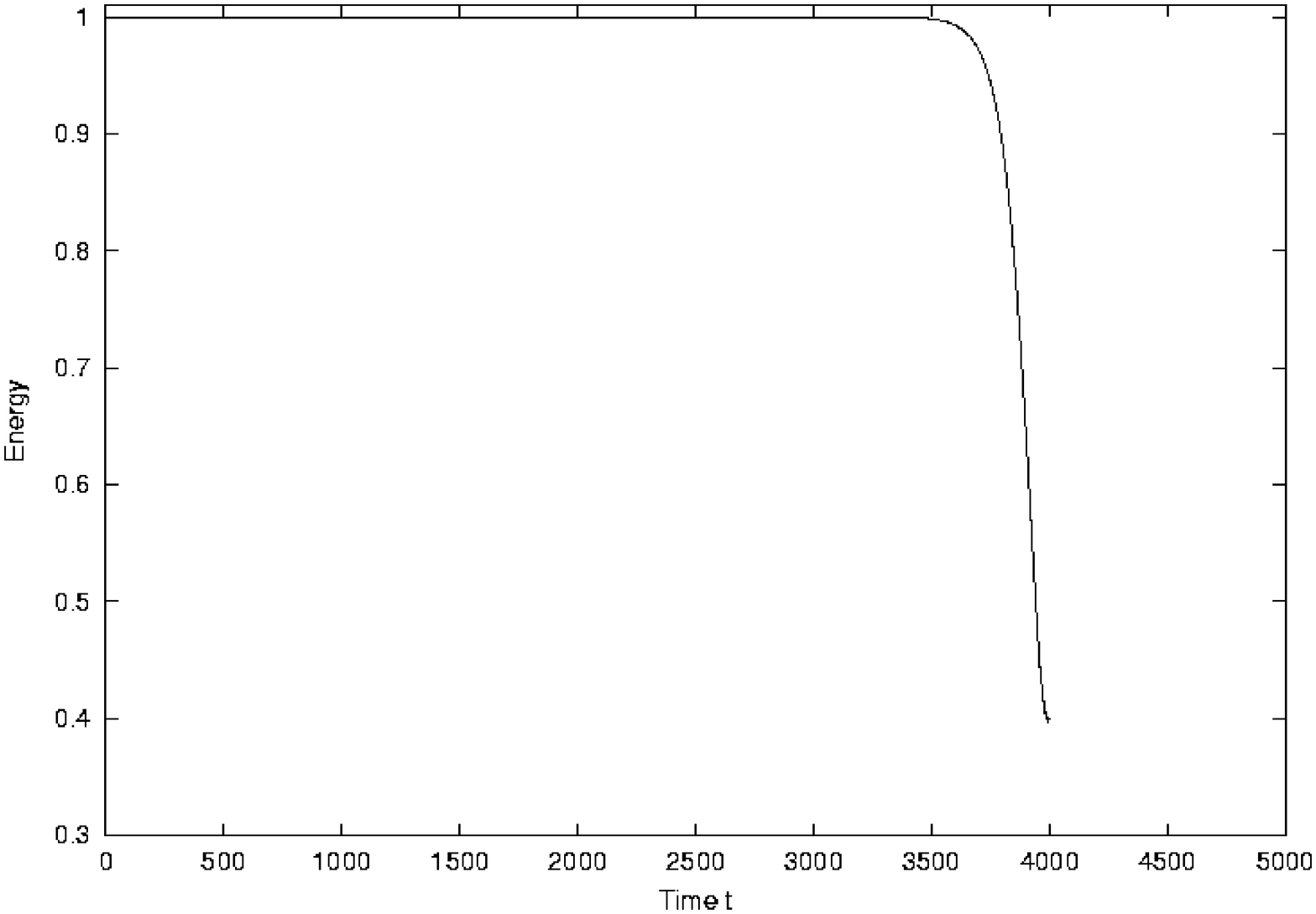}}
\caption{Energy of the mode $16$ vs $t$ for $\epsilon \mu = 0.005$ }
\label{fig:xt2}
\centerline{\includegraphics[angle=-90,width=.6\textwidth]{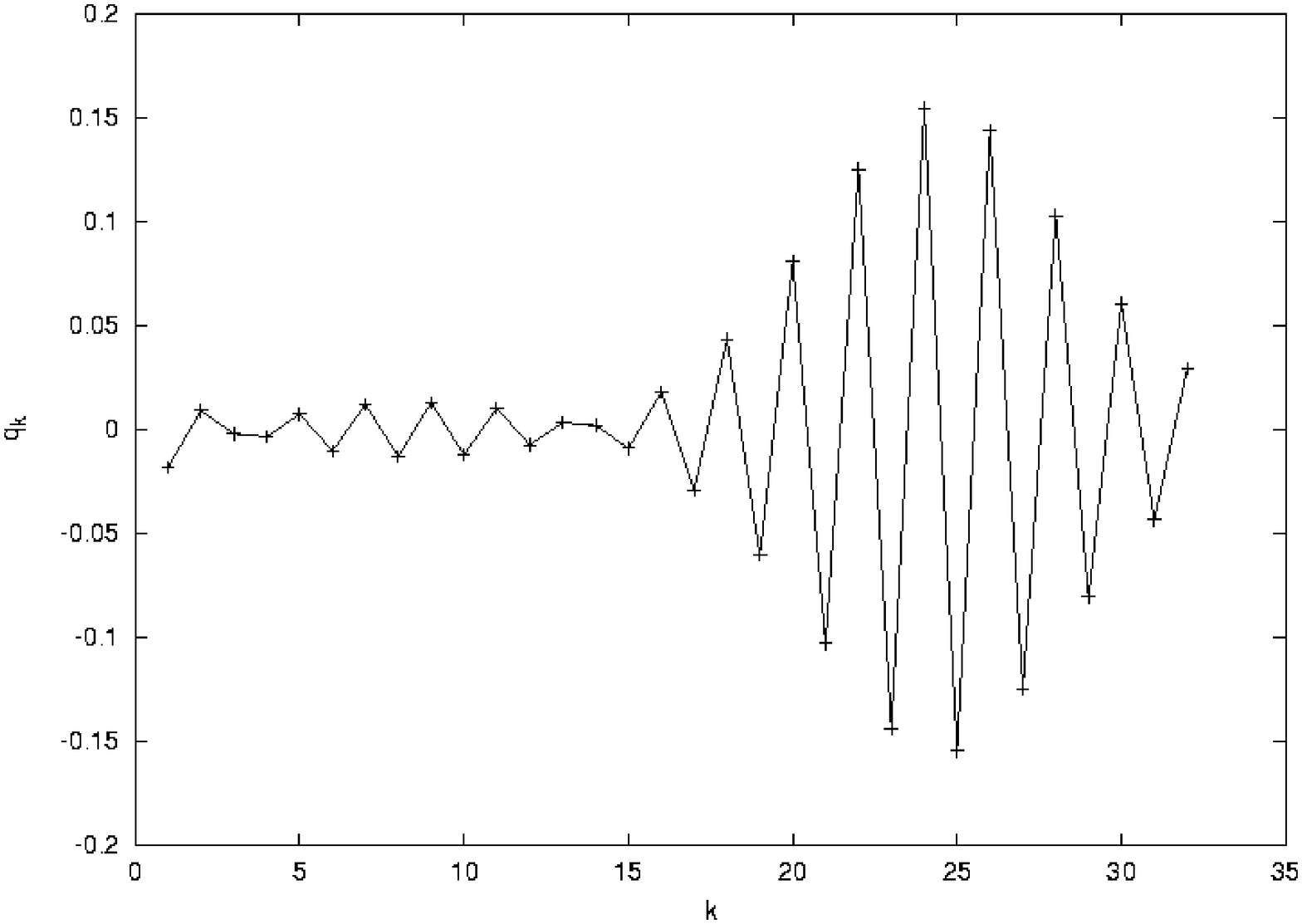}}
\caption{Values of the displacements of atoms in correspondence of the final value of energy in Fig. \ref{fig:xt2}}
\label{fig:xt1}
\end{figure}

\begin{figure}[htbp]
\centerline{\includegraphics[angle=-90,width=.6\textwidth]{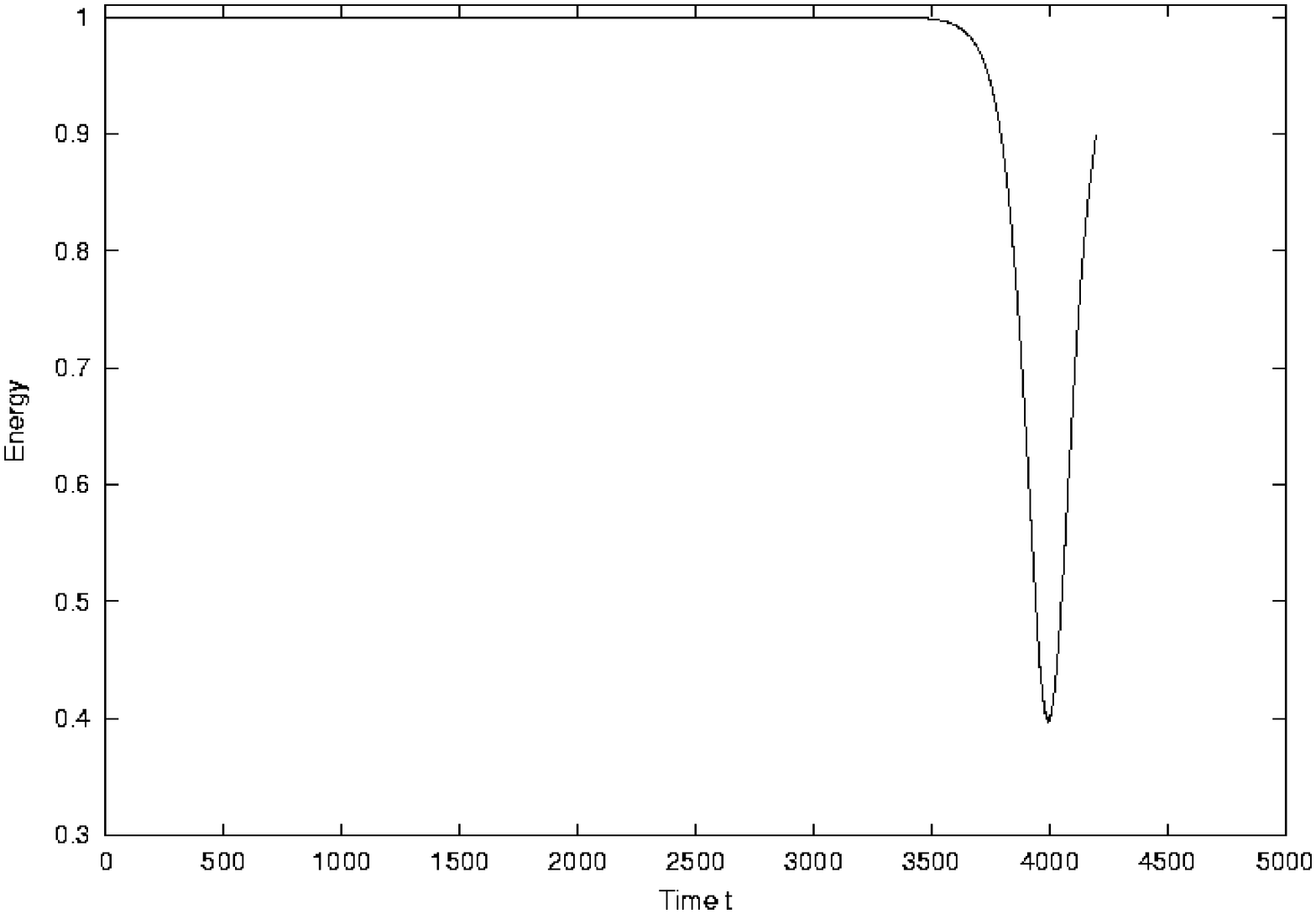}}
\caption{Energy of the mode $16$ vs $t$ for $\epsilon \mu = 0.005$ }
\label{fig:xt4}
\centerline{\includegraphics[angle=-90,width=.6\textwidth]{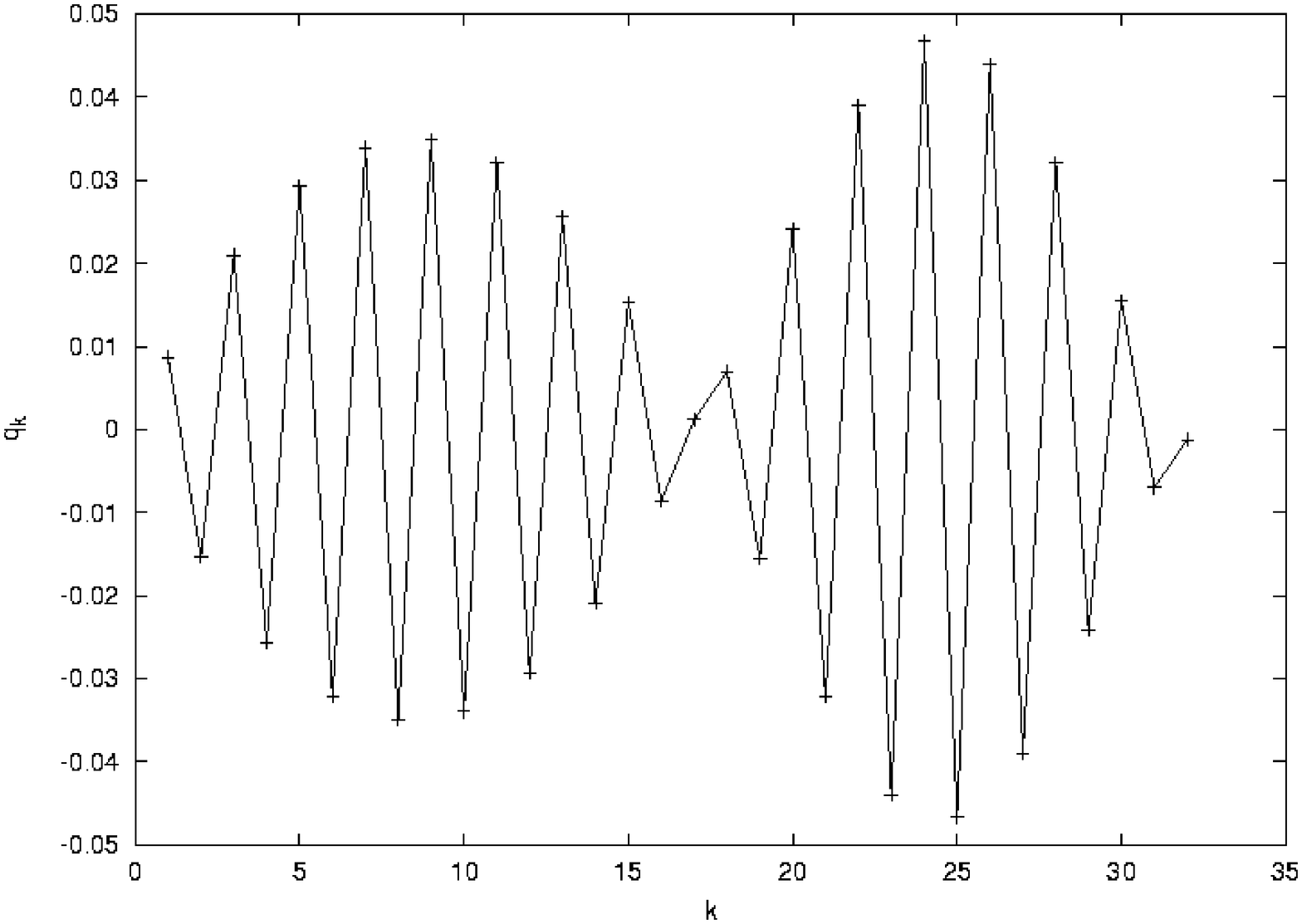}}
\caption{Values of the displacements of atoms in correspondence of the final value of energy in Fig. \ref{fig:xt4}}
\label{fig:xt3} 
\end{figure}

\begin{figure}[htbp]
\centerline{\includegraphics[angle=-90,width=.6\textwidth]{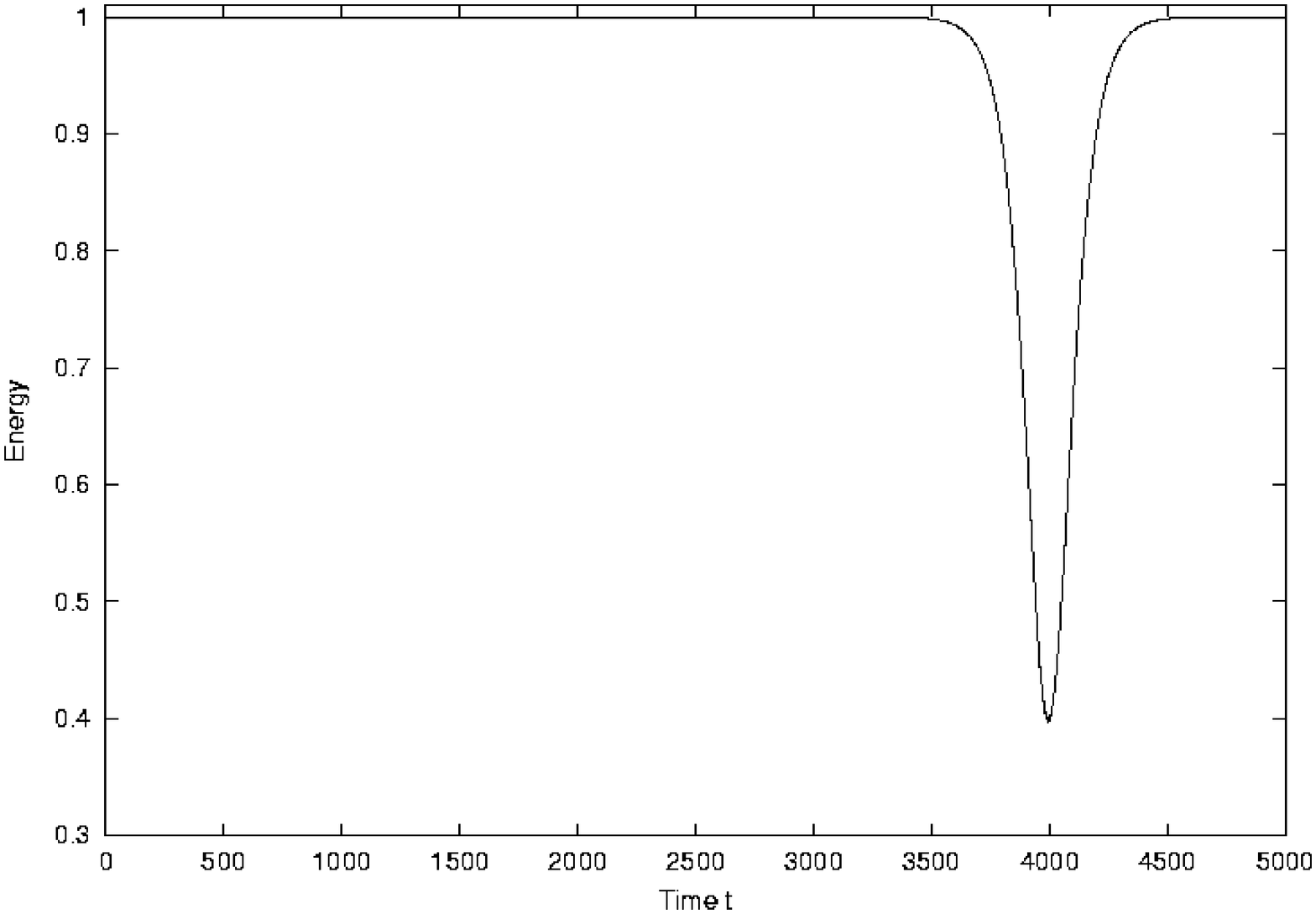}}
\caption{Energy of the mode $16$ vs $t$ for $\epsilon \mu = 0.005$ }
\label{fig:xt8}
\centerline{\includegraphics[angle=-90,width=.6\textwidth]{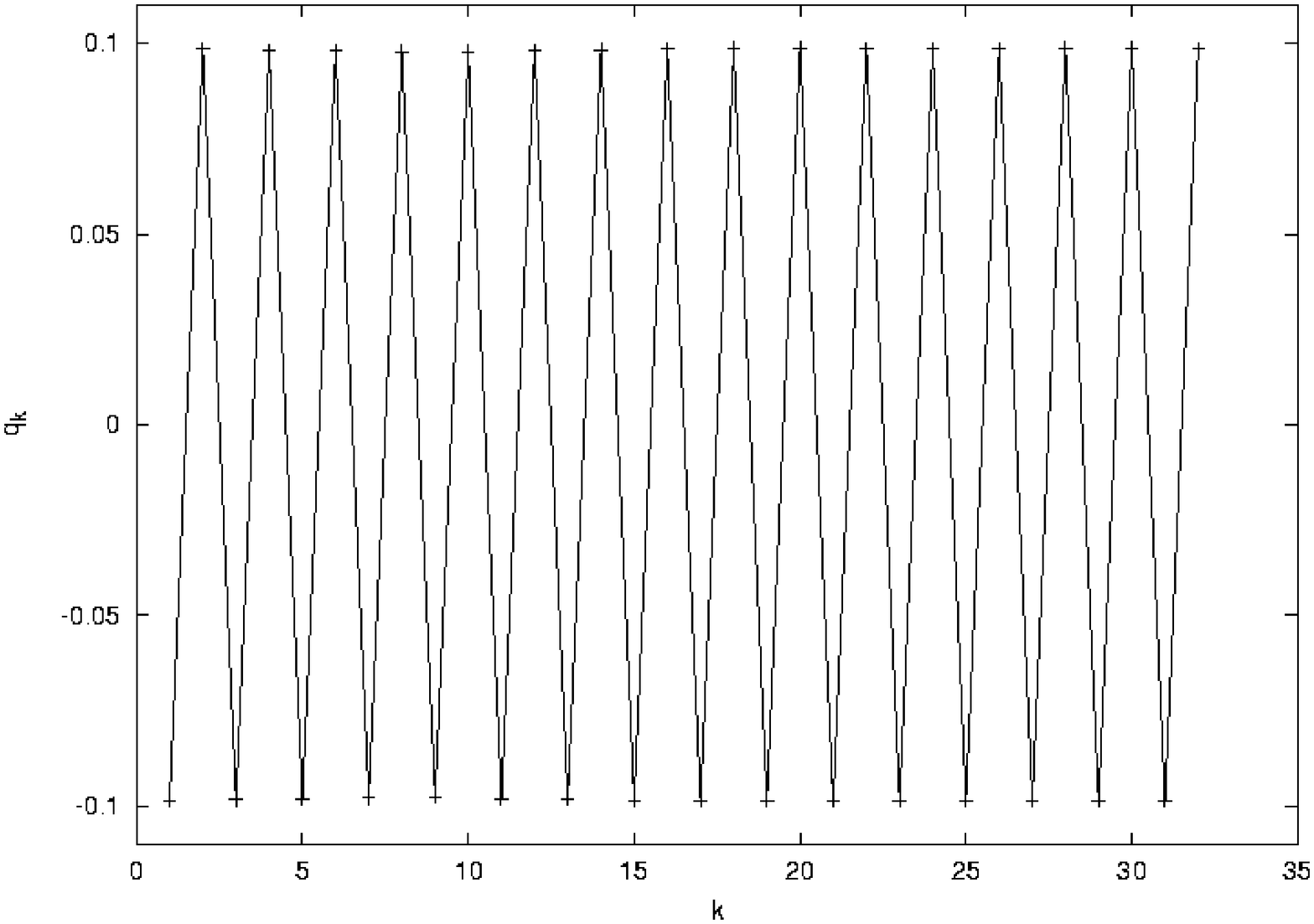}}
\caption{Values of the displacements of atoms in correspondence of the final value of energy in Fig. \ref{fig:xt8}}
\label{fig:xt7}
\end{figure}

In this context, and to obtain more insight in the behaviour of OMS, it is really very 
interesting to see what are the positions of particles in the chain, i.e. the spatial configuration of the chain, 
in correspondence of a well determined value of energy of the OMS. In figs \ref{fig:xt2} - \ref{fig:xt7} for 
$\epsilon \mu = 0.005$, the values of the coordinates of the 32 atoms, in correspondence of a particular value of 
the energy of mode $16$, are shown. To be clear, the representative points of the particles are joined by segments.

\bigskip  
\eject
\bigskip  
\eject
\bigskip 
\eject
As can be seen from the previous figures, the particle chain recovers its "simmetrical form" ($q_{k} + q_{k-1} = 0$) 
when the OMS recovers all its initial energy. 

\section{The case $N/2$ as a function of $N$}

To complete the analysis of the case $N/2$, in the next figures we show the behaviour of the energy of this mode 
as a function of time for $N = 26, 38, 52$ and $54$, for $\epsilon \mu = 0.005$. 

\begin{figure}[htbp]
\centering \begin{tabular}{cc}
\resizebox*{8 cm}{!}{\rotatebox{-90}{\includegraphics{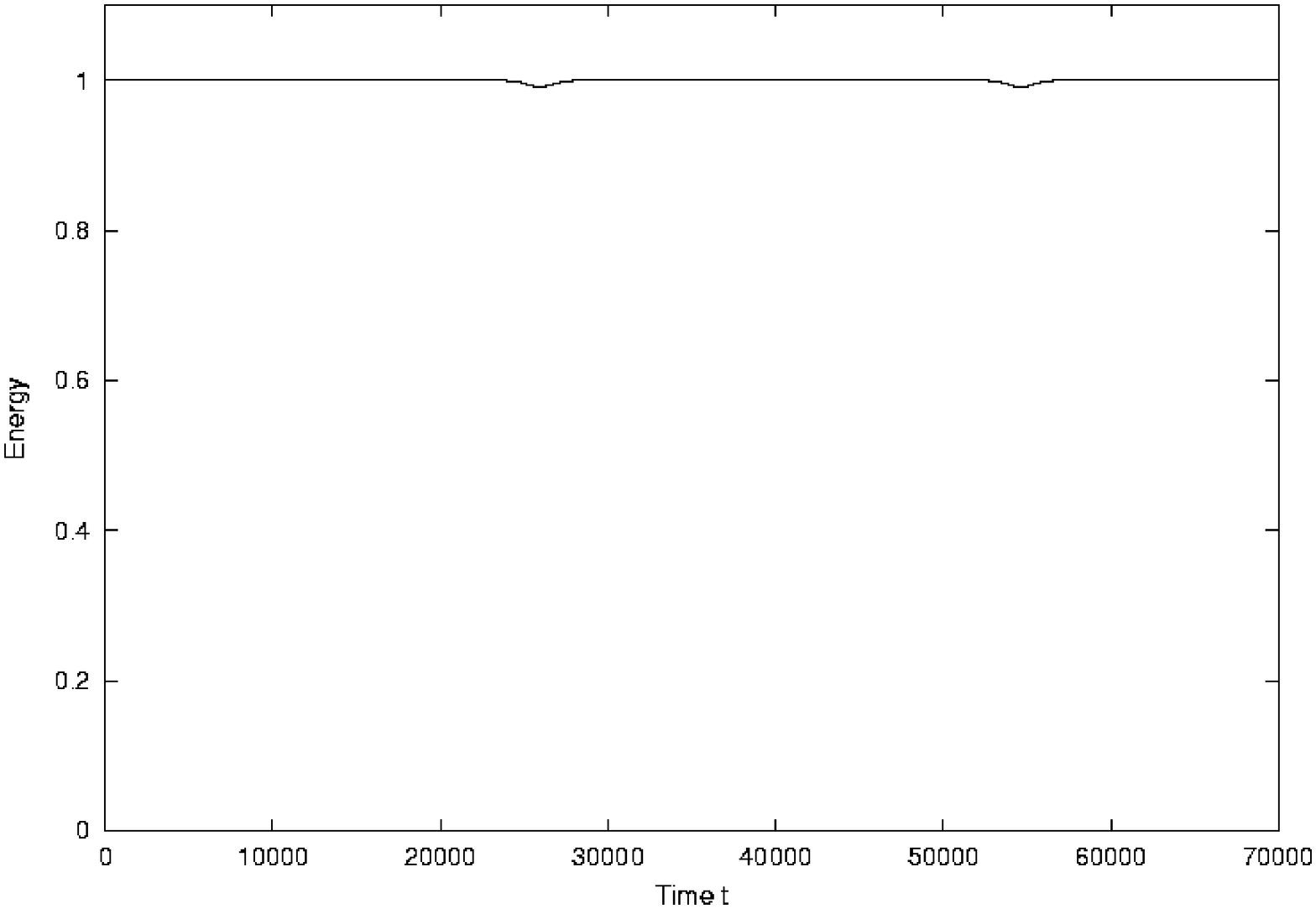}}} 
\resizebox*{8 cm}{!}{\rotatebox{-90}{\includegraphics{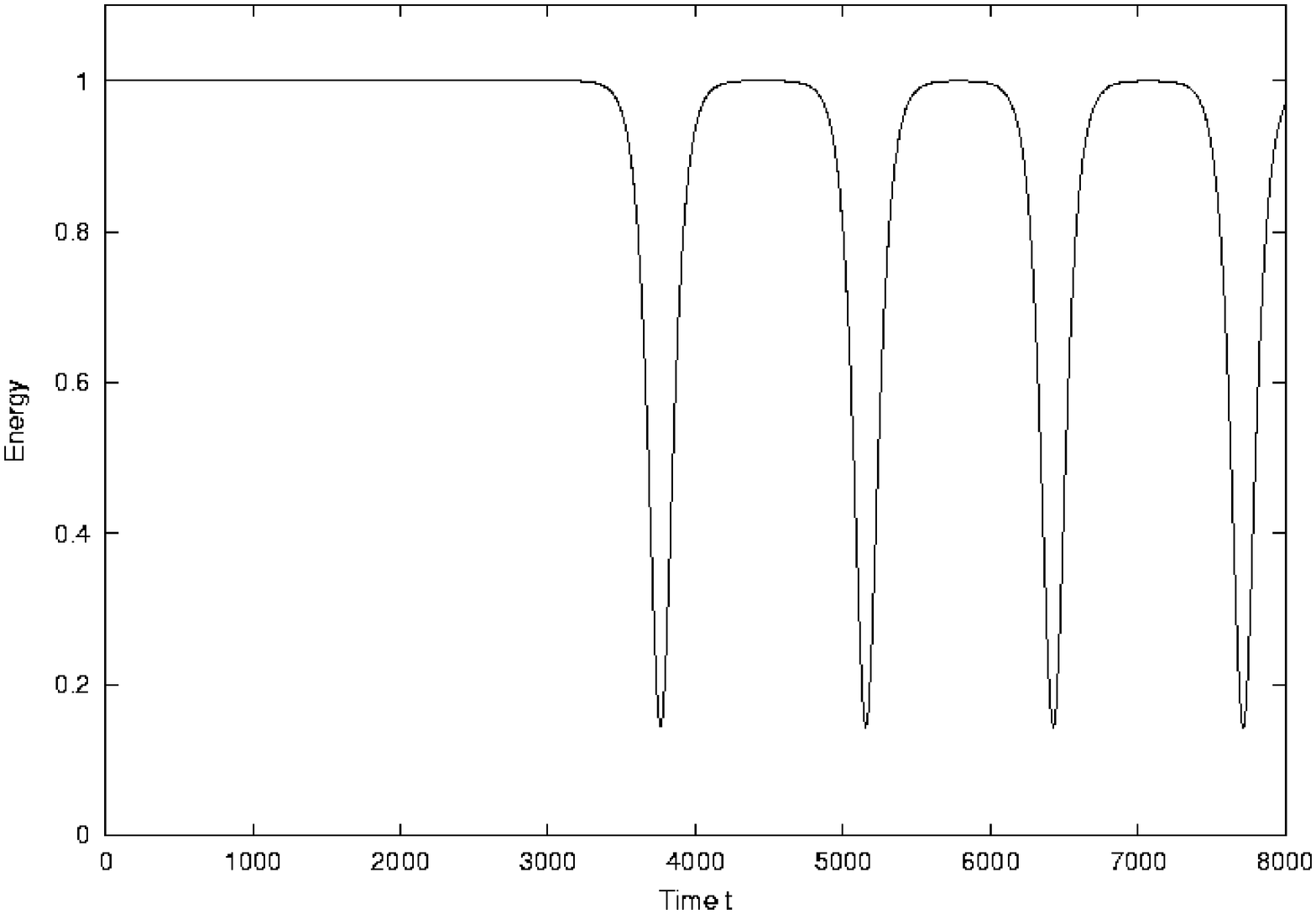}}}\\ 
\end{tabular}
\caption{Energy of the mode $N/2$ vs $t$ for $N = 26$ e $N = 38$.}
\end{figure}
\begin{figure}[htbp]
\centering \begin{tabular}{cc}
\resizebox*{8 cm}{!}{\rotatebox{-90}{\includegraphics{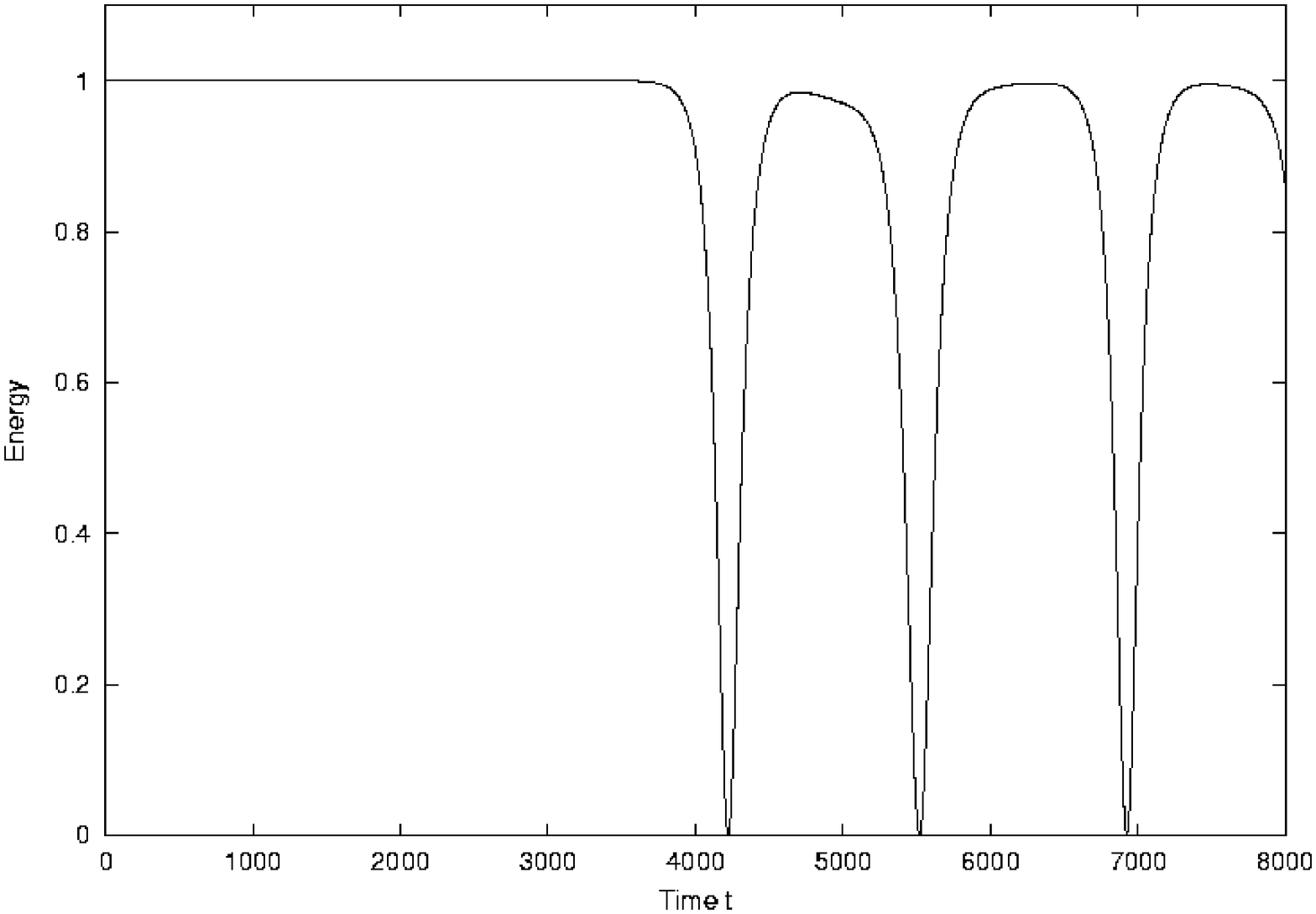}}} 
\resizebox*{8 cm}{!}{\rotatebox{-90}{\includegraphics{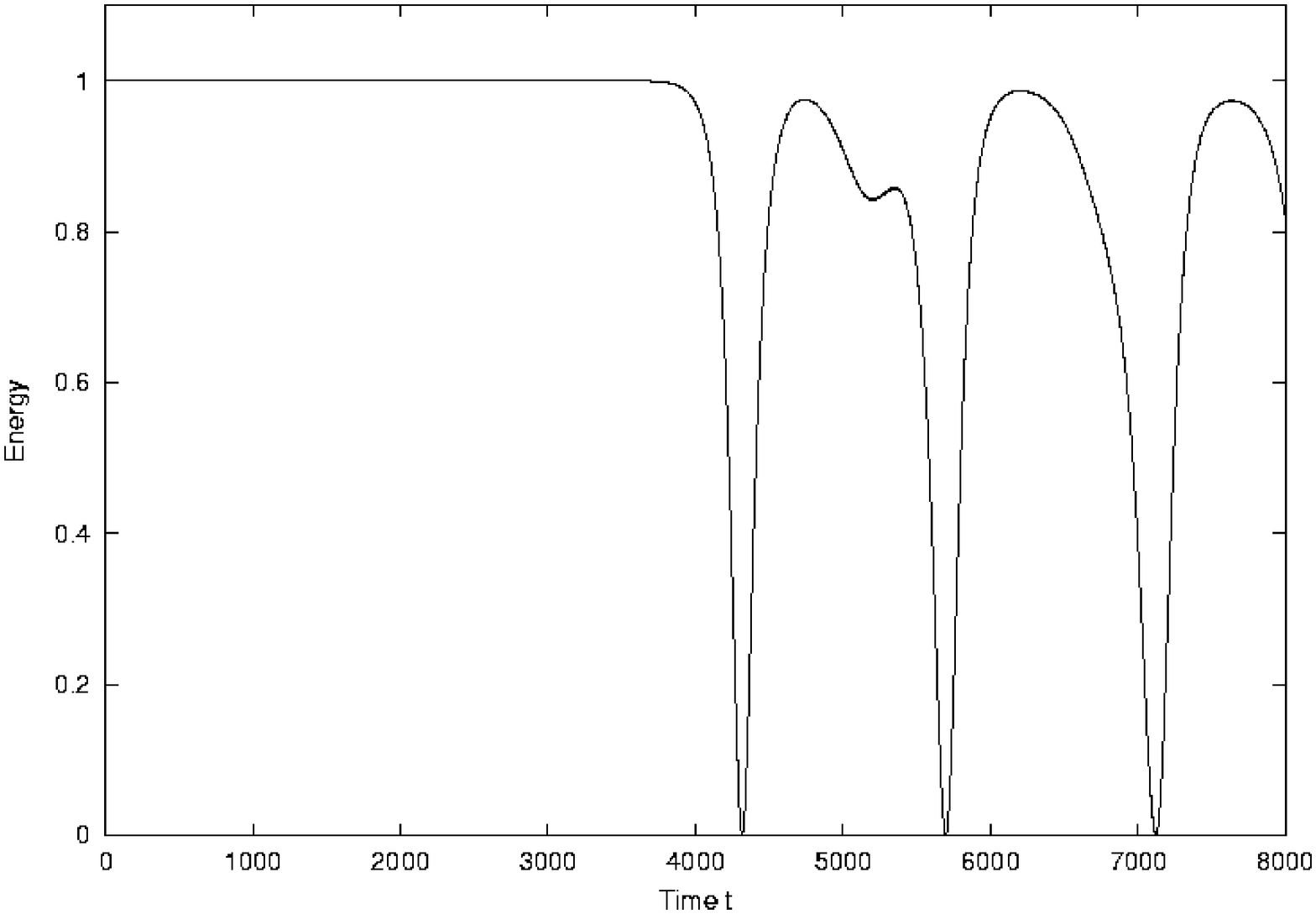}}}\\ 
\end{tabular}
\caption{Energy of the mode $N/2$ vs $t$ for $N = 52$ e $N = 54$.}
\end{figure}

\begin{figure}[htbp]
\centerline{\includegraphics[angle=-90,width=.7\textwidth]{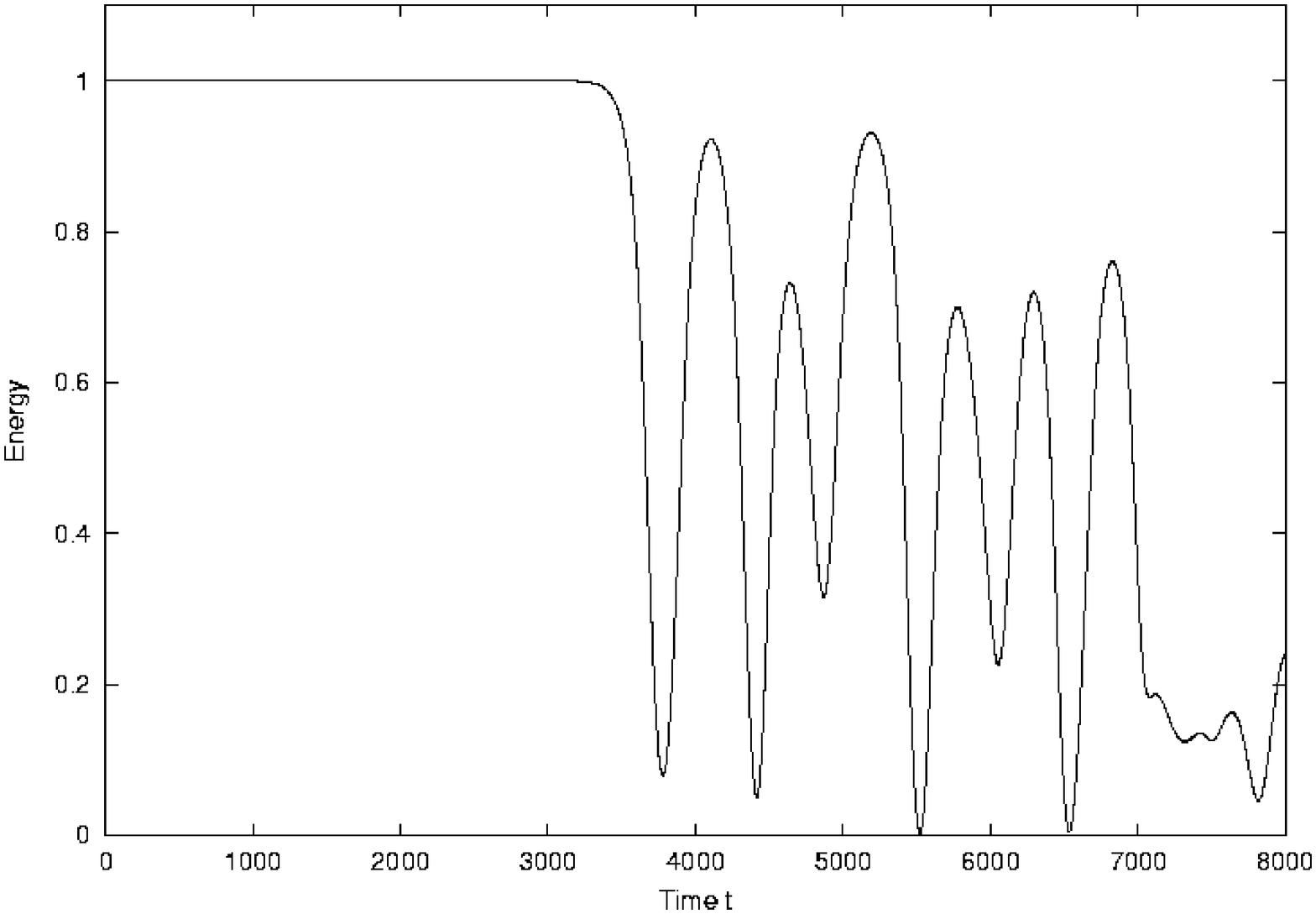}}
\caption{Energy of the mode $128$ for $N = 256$ and $\epsilon\mu = 0.005$ }
\label{fig:na256p}
\centerline{\includegraphics[angle=-90,width=.7\textwidth]{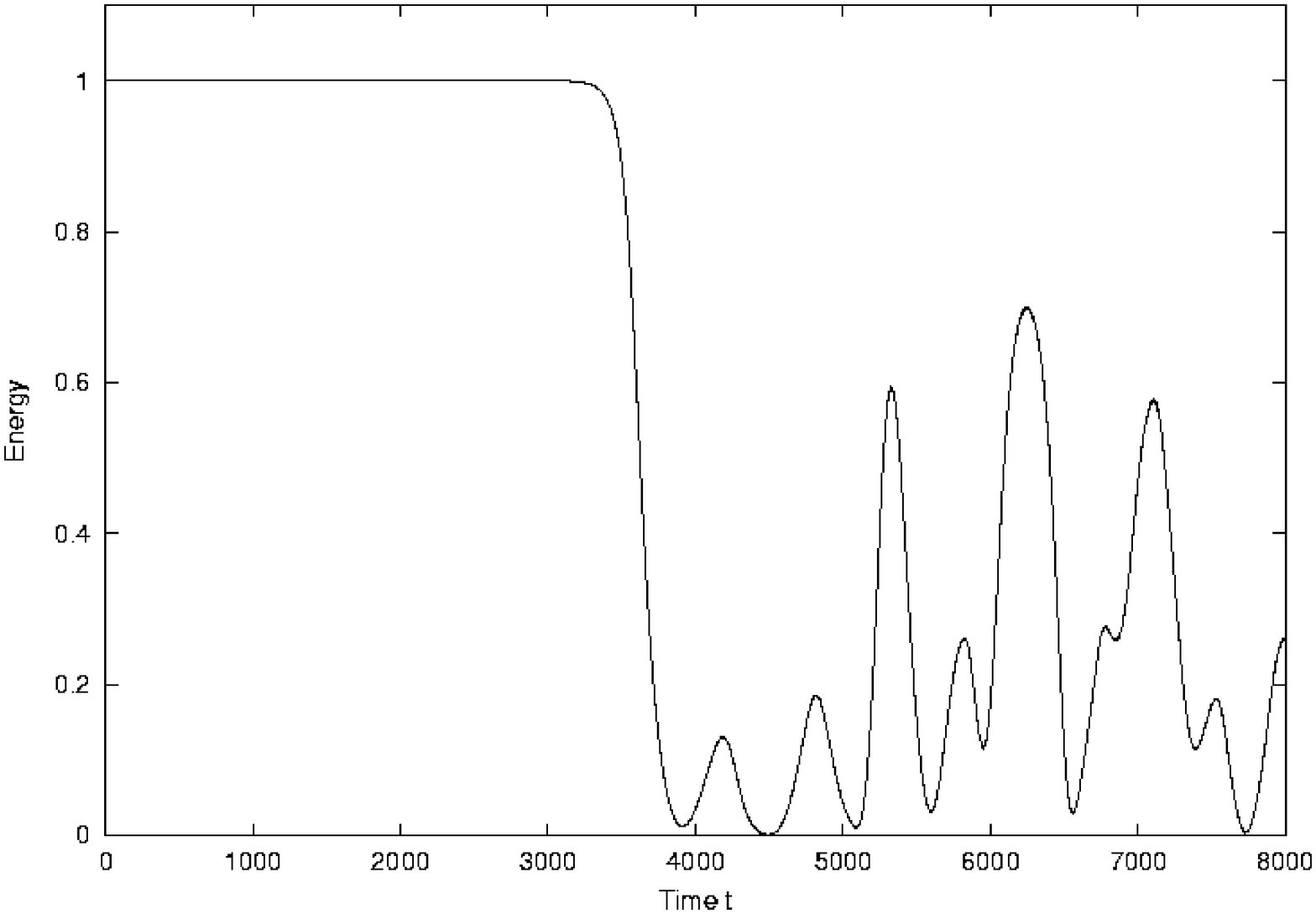}}
\caption{Energy of the mode $256$ for $N = 512$ and $\epsilon\mu = 0.005$}
\label{fig:na512p}
\end{figure}

From the previous figures it emerges that, increasing $N$, the mode $N/2$ tends to exchange all its energy. 
For $N = 52$, the mode $N/2$ periodically loses and recovers roughly all its energy. For $N > 52$, the recovery 
is not complete and moreover the curve of energy versus time becomes irregular. The irregularity and complexity 
of the curve increase with  $N$ and, for $N$ very large, the system becomes chaotic. This behaviour is evident 
in the  Figs. \ref{fig:na256p} and \ref{fig:na512p}, which show the behaviour of the energy  of the mode $N/2$, 
for $N = 256$ and $N = 512$.

\begin{figure}[htbp]
\centerline{\includegraphics[angle=-90,width=.9\textwidth]{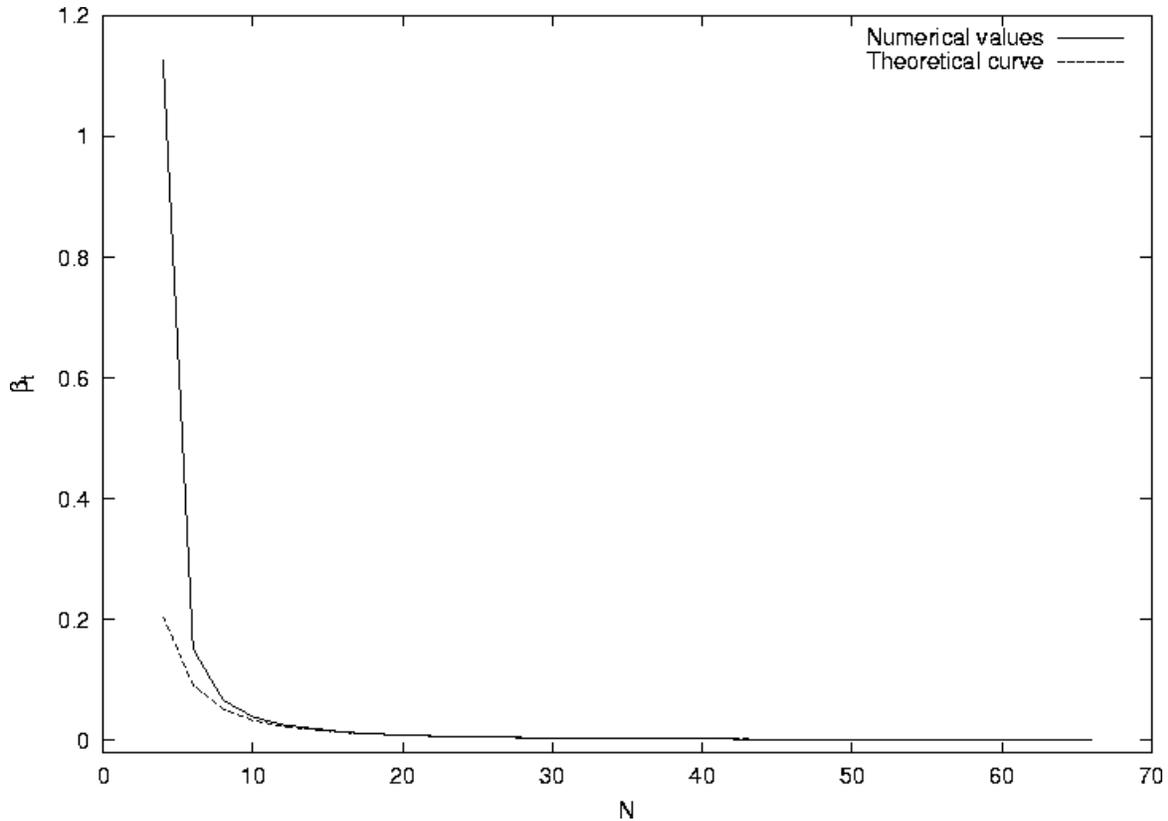}}
\caption{$\beta_{t}$ vs $N$ for the OMS $N/2$.}
\label{fig:soglia}
\end{figure}

\eject 
Let us now consider the formula (\ref{eq:beta}), obtained analytically, and let us see how we can obtain 
numerically the dependence on $N$ of the threshold energy density. From the analysis of the stability reported 
in \cite{bud} and in \cite{poggi}, we know that the first modes to be excited, as the energy increases and 
the mode $N/2$ becomes instable, are the modes $N/2 -1$ and $N/2 + 1$. Then for a value of $N$, starting 
from values of $\beta = \epsilon \mu$ very small, and integrating the motion equations for times fairly long, 
we increase the value of $\beta$ until the first pulse in the energy of the mode $N/2 -1$ or, equivalently, the 
first sudden change in the energy of the mode $N/2$, appears. We assume this value of $\beta$ as the value of  
$\beta_{t}$, the threshold energy density. In Fig. \ref{fig:soglia}, the numerical value of $\beta_{t}$, determined 
in this way, and the theoretical value of $\beta_{t}$ (formula (\ref{eq:beta})) are shown for $N = 4, 6, \ldots, 64$.    

\eject
\section{The cases $\frac{N}{4}$ and $\frac{3}{4} N$ }

\begin{figure}[htbp]
\centerline{\includegraphics[angle=-90,width=.4\textwidth]{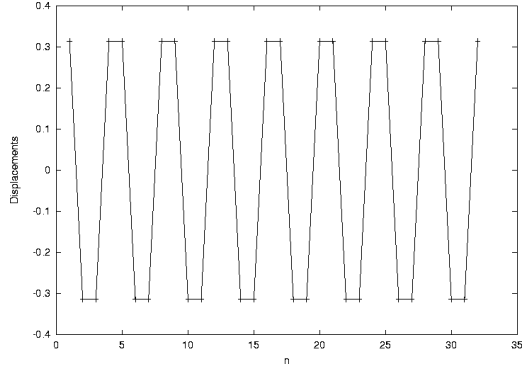}}
\caption{Initial configuration of the chain particles for the OMS  $N/4$ for $N = 32$ and 
$\epsilon\mu = 0.01$. The displacements from the equilibrium positions are shown vs $N$.}
\label{fig:confin}
\end{figure}

Following the same procedure used in the case $N/2$, we have calculated, numerically, the threshold energy 
density for the two nonlinear OMS's $N/4$ and $(3/4) N$ as  functions of $N$. We have found 
that $\beta_{t}$ is the same in the two cases. In the case $N/4$, the initial condition $Q_{N/4} \neq 0$, 
$P_{N/4} = 0$ corresponds, in the variables $q_{k}$, to the initial configuration shown in Fig. \ref{fig:confin}.

We have found that the nonlinear OMS's $N/4$ and $ (3/4) N$, for $N = 4$, $N = 8$ and $N = 12$, are stable 
in the limit of an integration time $T = 2 \times 10^4$.

\begin{figure}[htbp]
\centerline{\includegraphics[angle=-90,width=.6\textwidth]{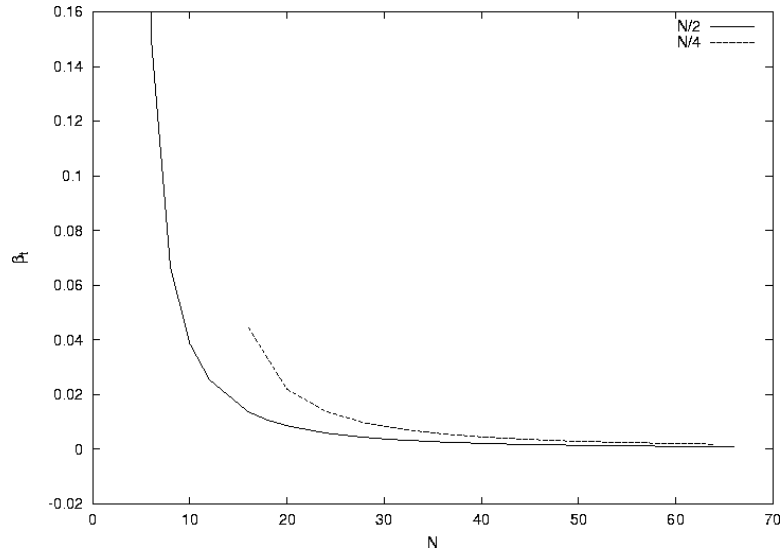}}
\caption{$\beta_{t}$ vs $N$ for the OMS's $N/2$ and $N/4$.}
\label{fig:ptss}
\end{figure}

The values of $\beta_{t}$, as a function of $N$, for the cases $N/2$ and $N/4$ are compared in 
Fig. \ref{fig:ptss}: \par

\section{The case $N/3$ and $\frac{2}{3}N$}
\begin{figure}[htbp]
\centerline{\includegraphics[angle=-90,width=.6\textwidth]{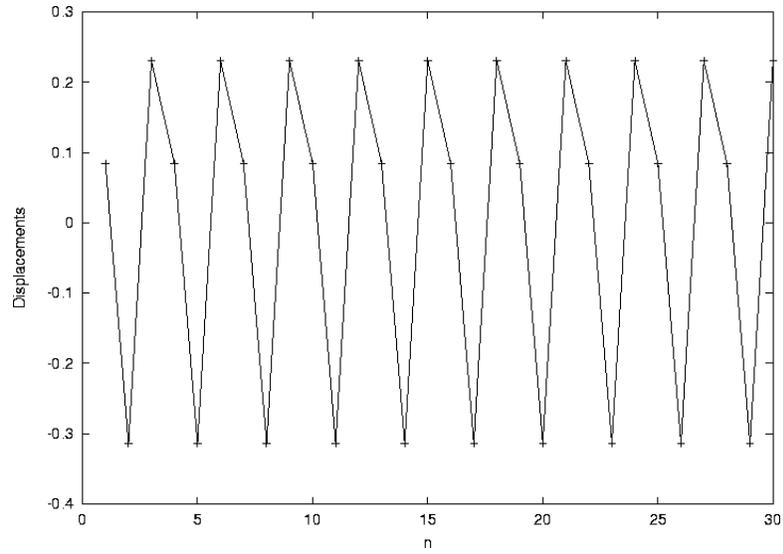}}
\caption{Initial configuration of the chain particles for the OMS $N/3$ for $N = 30$ and 
$\epsilon\mu = 0.008$. The displacements from the equilibrium positions, joined by segments, are shown vs $N$.}
\label{fig:confinn3}
\end{figure}

\begin{figure}[htbp]
\centerline{\includegraphics[angle=-90,width=.6\textwidth]{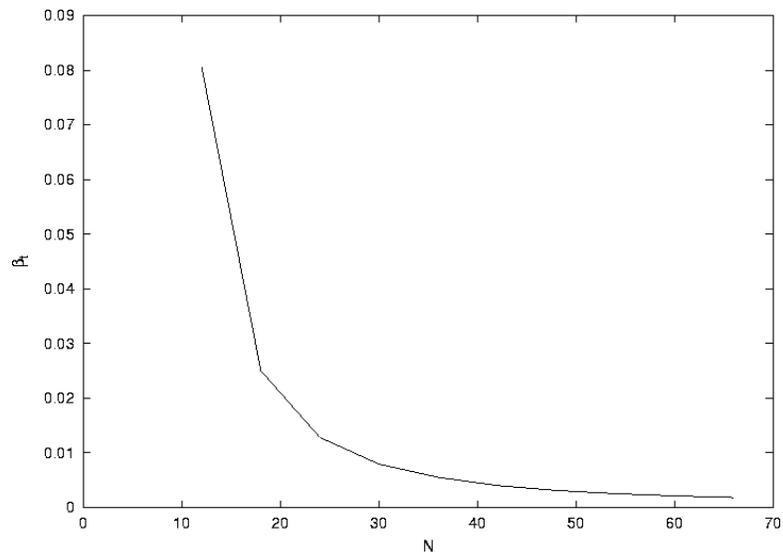}}
\caption{$\beta_{t}$ vs $N$ for the OMS's $N/3$ and $(2/3) N$}
\label{fig:thren323n}
\end{figure}

In this case, the initial condition: $Q_{N/3} \neq 0$, 
$P_{N/4} = 0$ corresponds, in the variables $q_{k}$, to the initial configuration shown in 
Fig. \ref{fig:confinn3}.
\eject
The behaviour of $\beta_{t}$, as a function of $N$, for the two cases, is given in the Fig. \ref{fig:thren323n}: \par
\bigskip
\eject 
In the following table, for some even value of $N$ between $4$ and $64$, the threshold $\beta_{t}$, for 
each OMS, is given. \par

\begin{tabular}{|l|l|l|l|l|l|r|} \hline\hline
      &        &         &         &        &        &           \\     
N     & $(\beta_{t})_{\frac{N}{4}}$~~~~~~~       & $(\beta_{t})_{\frac{N}{3}}$~~~~~~~   & $(\beta_{t})_{\frac{N}{2}}$~~
~~~~~& $(\beta_{t})_{\frac{1}{3}(\frac{\pi}{N})^{2}}$~~~~~~~ & $(\beta_{t})_{\frac{3}{4}N}$~~~~~~~   & 
$(\beta_{t})_{\frac{2}{3}N}$                 \\ 
      &        &         &         &         &        &          \\ \hline
~4    & stable &         & 1.1272 & 0.2056 & stable &          \\ \hline
~6    &        & stable  & 0.1514 & 0.0914 &        & stable   \\ \hline
~8    & stable &         & 0.0665 & 0.0514 & stable &          \\ \hline
10    &        &         & 0.0385 & 0.0329 &        &          \\ \hline
12    & stable & 0.0807  & 0.0254 & 0.0228 & stable & 0.0807   \\ \hline
16    & 0.0447 &         & 0.0136 & 0.0128 & 0.0447 &          \\ \hline  
18    &        & 0.0250  & 0.0106 & 0.0101 &        & 0.0250   \\ \hline
20    & 0.0219 &         & 0.0086 & 0.0082 & 0.0219 &          \\ \hline
24    & 0.0138 & 0.0128  & 0.0059 & 0.0057 & 0.0138 & 0.0128   \\ \hline
28    & 0.0096 &         & 0.0043 & 0.0042 & 0.0096 &          \\ \hline
30    &        & 0.0079  & 0.0037 & 0.0037 &        & 0.0079   \\ \hline
32    & 0.0072 &         & 0.0033 & 0.0032 & 0.0072 &          \\ \hline
36    & 0.0056 & 0.0055  & 0.0026 & 0.0025 & 0.0056 & 0.0055   \\ \hline
40    & 0.0045 &         & 0.0021 & 0.0021 & 0.0045 &          \\ \hline 
42    &        & 0.0040  & 0.0019 & 0.0019 &        & 0.0040   \\ \hline
44    & 0.0037 &         & 0.0018 & 0.0017 & 0.0037 &          \\ \hline
48    & 0.0031 & 0.0031  & 0.0015 & 0.0014 & 0.0031 & 0.0031   \\ \hline 
52    & 0.0027 &         & 0.0013 & 0.0012 & 0.0027 &          \\ \hline
54    &        & 0.0025  & 0.0012 & 0.0011 &        & 0.0025   \\ \hline
56    & 0.0023 &         & 0.0011 & 0.0010 & 0.0023 &          \\ \hline
60    & 0.0021 & 0.0021  & 0.0010 & 0.0009 & 0.0021 & 0.0021   \\ \hline
64    & 0.0018 &         & 0.0009 & 0.0008 & 0.0018 &          \\ \hline
66    &        & 0.0018  & 0.0009 & 0.0008 &        & 0.0018   \\ \hline\hline 
\end{tabular}
\par
The error is of one unit on the last digit. \par
\begin{figure}[htbp]
\centerline{\includegraphics[angle=-90,width=.6\textwidth]{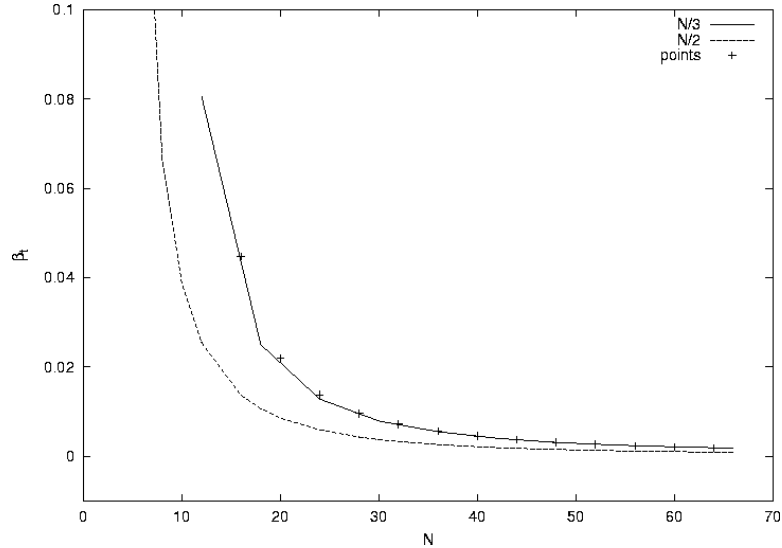}}
\caption{$\beta_{t}$ vs $N$ for the cases $N/2$, $N/3$ and $N/4$.}
\label{fig:n2n3n4}
\end{figure}

In Fig. \ref{fig:n2n3n4} the behaviour of $\beta_{t}$ as a function of $N$ is given for the  cases $N/2$, 
$N/3$ and $N/4$. The first two values of $\beta_{t}$ of the case $N/2$ have been left out for clarity. \par
\eject
\section{\bf Discussion} 
We have numerically analyzed the stability of the OMS's in the Fermi-Pasta-Ulam system. Previously, 
only for the case $N/2$ were available theoretical analysis of the stability  and approximate estimates of the 
stability threshold for large values of $N$. Our method is based on the numerical integration of full nonlinear 
differential equations of motion for the particles of the chain. The initial conditions for the hamiltonian variables 
$q_{k}$ and $p_{k}$ are such that only a particular nonlinear one-mode is initially excited. No a priori initial 
perturbation of the analytical solution is introduced in the numerical algorithm, the only perturbation being 
that generated by computational errors in numerical integration. Then the method studies the stability of the OMS's 
against the numerical errors introduced by the integration algorithm. We have widely analysed the case $N/2$, 
which in some sense works as a test, since, for this case, theoretical results are available. 

\begin{figure}[htbp]
\centerline{\includegraphics[angle=-90,width=.8\textwidth]{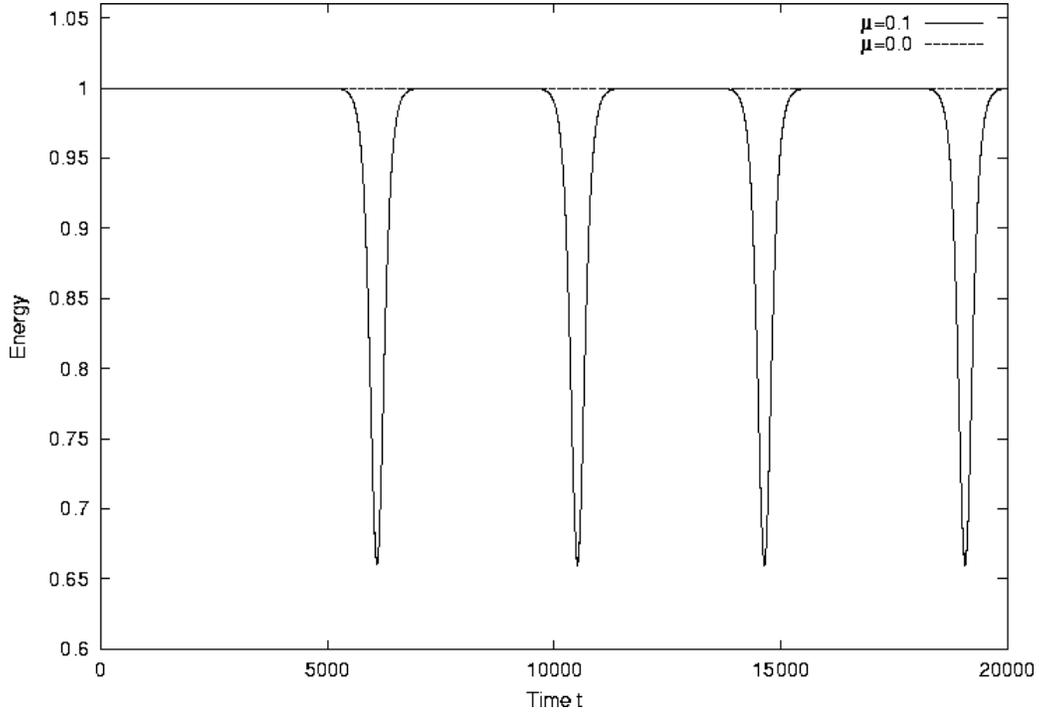}}
\caption{Energy as a function of time for $\epsilon = 0.04$ in the linear case $\mu = 0$, and 
$\mu = 0.1$.}
\label{fig:confront}
\end{figure}

\begin{figure}[htbp]
\centerline{\includegraphics[angle=-90,width=.6\textwidth]{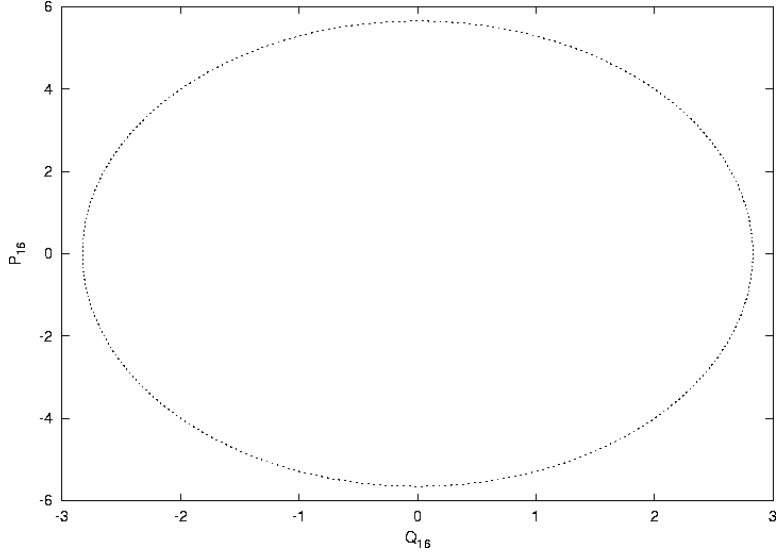}}
\caption{Orbit in the plane $Q_{16},P_{16}$ for the linear case ($\mu = 0$), with $\epsilon = 0.5$.}
\label{fig:confrontl}
\end{figure}

\begin{figure}[htbp]
\centerline{\includegraphics[angle=-90,width=.6\textwidth]{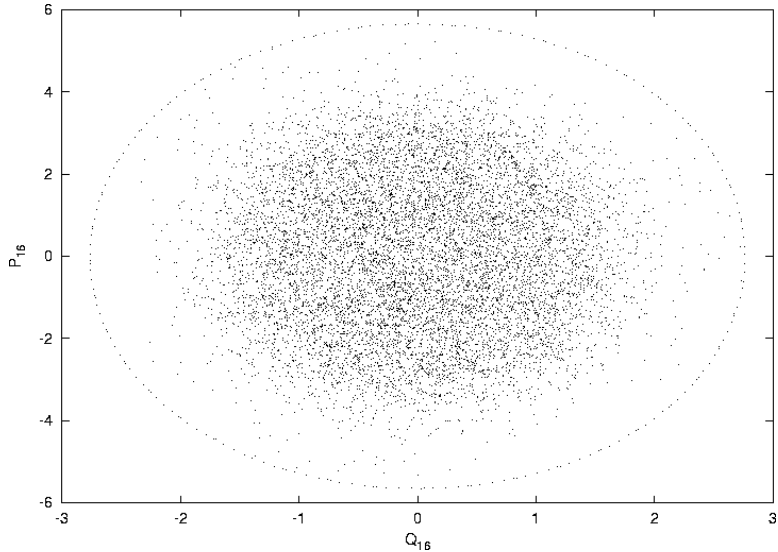}}
\caption{Orbit in the plane $Q_{16},P_{16}$ for the nonlinear case $\epsilon = 0.5$,  $\mu = 0.1$.}
\label{fig:confrontnl}
\end{figure}

We remark that the OMS's are nonlinear analytical solutions of the complete Fermi-Pasta-Ulam system, 
with linear and nonlinear terms in the hamiltonian, so their stability can't be discussed in terms of KAM theorem. 
The stability of linear modes and nonlinear OMS's, against computational errors, are completely 
disjoint in the sense that a linear mode, initially excited, is stable for very long integration times, if the 
parameter $\mu$ is set equal to zero in the hamiltonian. Thus the linear modes, excited during the evolution of 
a nonlinear OMS, are triggered by the instability of this nonlinear mode and then only indirectly 
by the computational errors. This is evident in Fig. \ref{fig:confront}, where the behaviour of energy vs time of 
the mode $N/2$ for $N = 32$, $\mu = 0.1$ and $\epsilon = 0.04$ is compared with the behaviour of the energy in 
the linear case with the same value of energy density $\epsilon = 0.04$ and $\mu = 0$. 
\eject
As can be seen from this figure, the linear case is stable with respect to the computational errors of the numerical 
algorithm. This different behaviour of the linear mode and of the nonlinear OMS is much more evident if we compare, 
for the same integration time, the orbit in the plane $Q_{N/2}, P_{N/2}$, for the same value of $\epsilon = 0.5$ 
and and the two values of $\mu = 0.0$, for the linear case, and $\mu = 0.5$ which corresponds, in the nonlinear 
case, to a value of $\beta$ very much larger than the threshold value $\beta_{t} = 0.0033$. This different behaviour 
is shown in Fig. \ref{fig:confrontl} and in Fig. \ref{fig:confrontnl}. \par
\vfill\eject
\bigskip
Let us first consider the case $N/2$. From the inspection of Figs. \ref{fi:qp01} - \ref{fi:nm4}, three regimes can 
be observed, varying the product $\beta = \epsilon\mu$. For values of $\beta$ very small, well below the 
threshold energy density, the nonlinear OMS $N/2$ is stable and $Q_{N/2}$ is a periodic function with 
the same amplitude and the same period of the analytical solution (\ref{eq:cn}). For values of $\beta$ above and 
near the threshold, the situation is very different. The period of oscillation is equal to the period of the 
analytical solution, and for very long times, the representative point moves in the plane $Q_{N/2}, P_{N/2}$ on 
a one dimensional closed curve, as for very small values of $\epsilon \mu$; but now, periodically and for short 
times, the amplitude of oscillation varies, due to a decrease of modal energy, and the representative point 
of the system moves on an open curve  which tends periodically to shrink. For $\epsilon \mu =0.1$, well above the 
threshold energy density, we observe a chaotic behaviour. 
 
In the first regime, where $\beta < \beta_{t}$, during the whole integration time, the OMS appears stable. 

As soon as $\beta$ exceeds $\beta_{t}$, after an initial time interval, depending on the precision of numerical 
calculations, during which the OMS is stable, the amplification of errors excites the  first modes which become 
unstable, namely the modes $N/2 -1$ and $N/2 + 1$. Because of the nonlinear coupling between the modes, the 
excitation of this two modes triggers all the other linear modes. The characteristics of this second regime, when 
the parameter $\beta$ grows, are the increasing exchange of energy between the nonlinear mode and the other linear 
modes and the periodic recovery of energy of the nonlinear mode. This regime continues as 
long as the periodic exchange of energy is complete. Further increase of $\beta$, distorts the profile of the 
curve of energy of the nonlinear mode, as a function of time, and subsequently the system becomes chaotic.

\begin{figure}[htbp]
\centerline{\includegraphics[angle=-90,width=.6\textwidth]{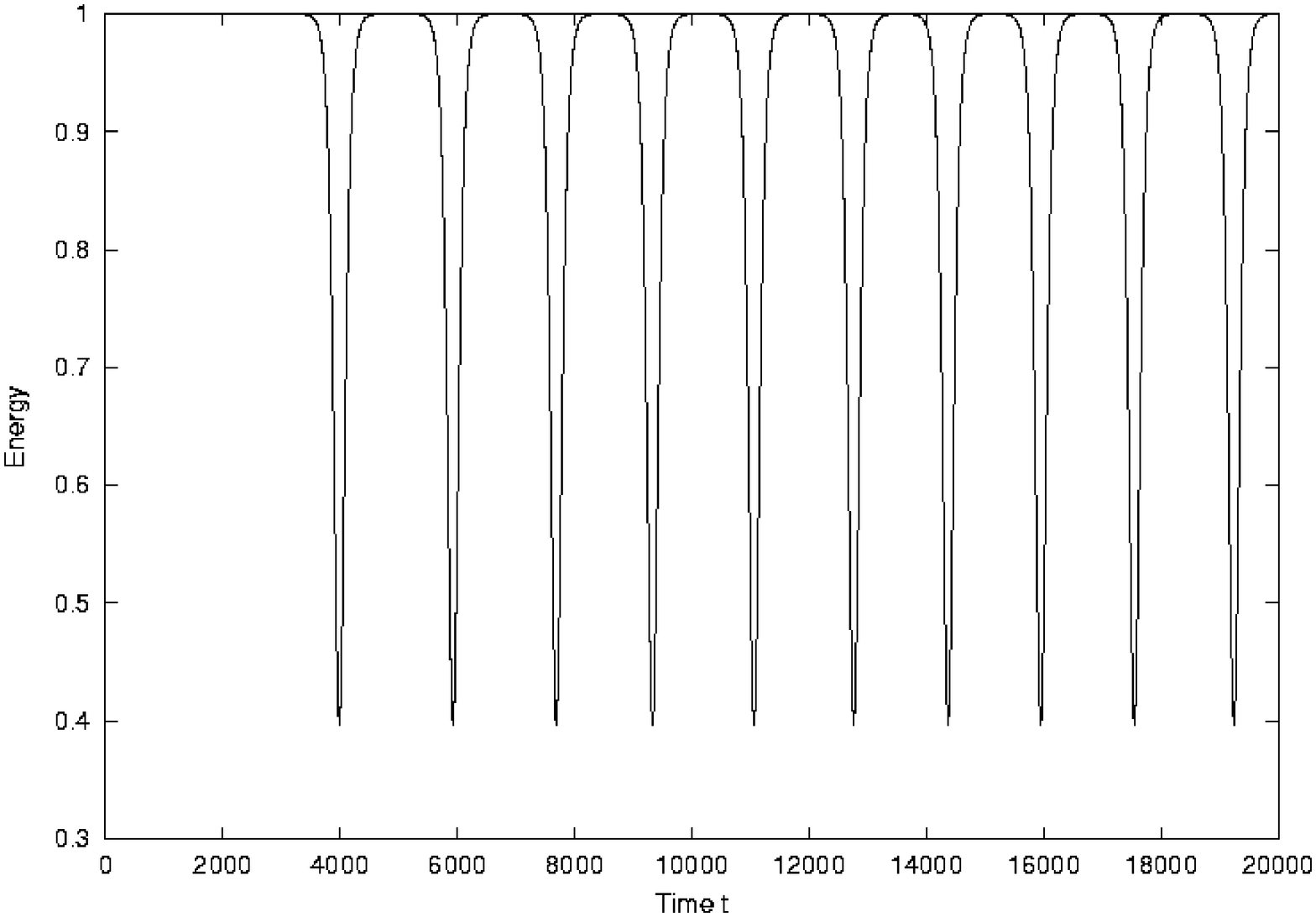}}
\caption{Energy of the mode 16 vs t for $\epsilon \mu = 0.005$.}
\label{fig:ener1s}
\centerline{\includegraphics[angle=-90,width=.6\textwidth]{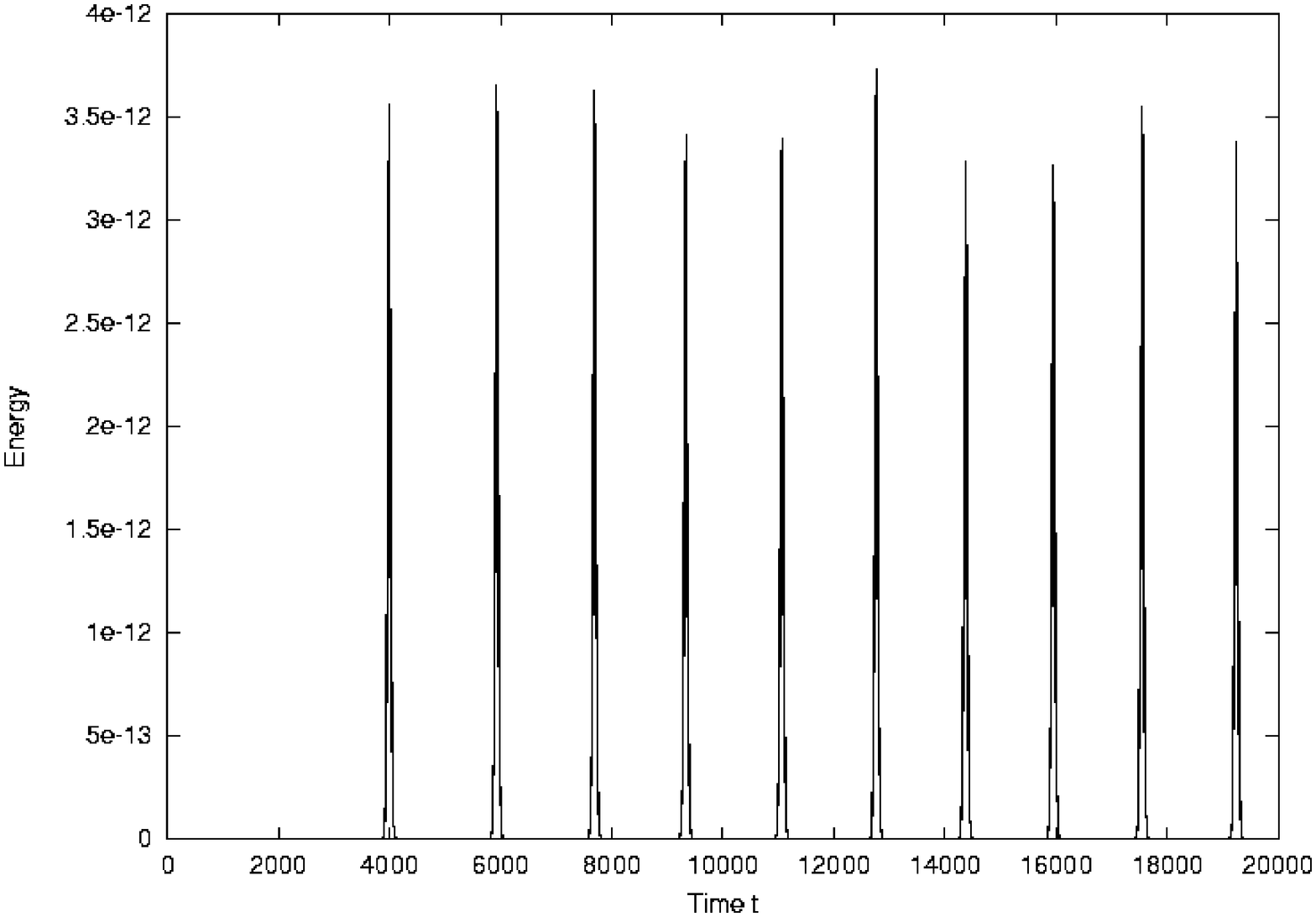}}
\caption{Energy of the mode 6 vs t for $\epsilon \mu = 0.005$.}
\label{fig:enerq6}
\end{figure}

We have seen that, for large values of $N$, formula (\ref{eq:beta}) gives a good estimate of the instability 
threshold, so we could try to explain the "intermittent behaviour" of the second regime in the framework of the 
theory developed in \cite{poggi}. As we have seen in section (4), modes with $\rho < 1/3$, are always stable in 
the linear approximation, for any energy density of the mode $n = N/2$, so that, perturbation of this mode, 
involving only modes with $\rho < 1/3$, never leads to instability. Let us consider for example the case $N = 32$. 
We have, from (\ref{eq:rho}), that the modes with $r < 7$ are always stable in the linear approximation, for any 
energy density of mode $n = N/2 = 16$. These modes can grow only if they are triggered  by the interaction with 
other unstable modes. As is pointed out in \cite{poggi}, this kind of interaction is neglected in the 
linear approximation and comes into play only when unstable modes have grown and the linearized theory is no 
longer valid. This indirect triggering is evident in the Figs. \ref{fig:ener1s} - \ref{fig:enerq6}, where the 
time evolutions of the modes $16$ and $6$ are compared.
No triggering of the mode $6$ exists, if $\beta < \beta_{t}$.

For $\beta > \beta_{t}$, in the short time intervals, during which the nonlinear one-mode exchanges energy with 
the other modes, the time derivative of the energy of the OMS is not zero and the equation (\ref{eq:q2p}) and the 
formula (\ref{eq:en}) are no longer valid. During these time intervals, a strong coupling exists between the 
mode $N/2$ and the two adiacent modes and, indirectly, with all the other linear modes. However, the 
exact mechanism by which the nonlinear one-mode loses and recovers periodically energy is not clear. We try now to 
give an explanation of this mechanism in terms of Floquet's theorem for parametric oscillators \cite{landau}, \
\cite{fame}. Let us consider the equation:
\begin{equation}
\frac{d x^{2}}{d t^{2}} + f(t) x = 0,
\label{eq:ma}
\end{equation}

\noindent where $f(t)$ is a periodic function of time with period $T$. The Floquet's theorem means that the 
solution of (\ref{eq:ma}) can be written in the form:
\begin{equation}
x_{r}(t) = \mu_{r}^{t/T} X_{r}(t),
\label{eq:sol}
\end{equation}

\noindent where $r = 1$ or $2$, $\mu_{1}\mu_{2} =1$ and $X_{r}(t + T) = \pm X_{r}(t)$. The appropriate sign in 
front of $X_{r}(t)$ is determined by the particular form of $f(t)$. With the minus sign one has  
$X_{r}(t + 2T) = X_{r}(t)$. Solution (\ref{eq:sol}) means also that
\begin{equation}
x_{r}(t+T) = \mu_{r} x_{r}(t).
\label{eq:per}
\end{equation}

\noindent Thus the values of $x_{r}(t)$, in successive cycles, increases by a factor $\mu$ when the time interval 
between the observations is equal to the period of $f(t)$. Concerning the stability of the oscillator, we can  
consider three separate cases (\cite{pipes}, \cite{yorke}). 

The first case corresponds to  real values for both $\mu_{1}$ and $\mu_{2}$. Since $\mu_{1} \mu_{2} = 1$, one of 
them (e.g., $\mu_{1}$) is greater than unity. After $N - 1$ cycles of oscillations, the magnitude of the 
displacement $\mid x_{1}(0)\mid$ grows to the value ${\mid \mu_{1} \mid}^{N} \mid x_{1}(0) \mid$ which exceeds 
$\mid x_{1}(0)\mid$: the displacement therefore diverges exponentially with $N$ and the oscillations are unstable. 
The familiar case, known as "parametric resonance", occurs when $\mid\mu_{1}\mid > 1$ and the period of the 
Floquet function $X(t)$ is $2 T$, where $T$ is the period of $f(t)$.  

The second case corresponds to $\mid\mu_{1}\mid = \mid\mu_{2}\mid = 1$. After $N -1$ cycles of oscillation the 
size of the displacement  is equal to the initial value, i.e., the oscillations are stable.

In the third case, $\mu_{1}^{2} = \mu_{2}^2 = 1$, so that $\mu_{1}$ and $\mu_{2}$ have the same value and the case 
is "degenerate". In this case one can show that the degenerate oscillator is stable or unstable with 
displacement growing linearly with $N$. 

In the matrix theory of the stability of eq. (\ref{eq:ma}),  $\mu_{1}$ and $\mu_{2}$ are roots of a quadratic 
equation. If we define the displacement and velocity at the beginning of the $n$-th cycle of the function $f(t)$ as 
$x_{n}$ and $v_{n}$, respectively, the corresponding variables at the beginning of the next cycle of $f(t)$ are 
related to the vector $P(n) = (x_{n}, v_{n})$ by two linear equations which can be written as the matrix equation

\begin{equation}
P(n + 1) = M(T) P(n).
\label{eq:pmp}
\end{equation} 

\noindent Thus the quadratic equation for $\mu$ is

\begin{equation}
\mu^{2} - Tr[M(t)] \mu + Det[M(T)] = 0, 
\label{eq:tr}
\end{equation}

\noindent where $Tr[M(T)]$ ~and~ $Det[M(T)]$ are the  trace and determinant of the $2 \times 2$ matrix ~$M(T)$ 
~and~ $Det[M(T)] = \mu_{1} \mu_{2} = 1$.

The matrix method is particularly useful  in determining the stability of the solutions. Suppose that the graph of 
$f(t)$ vs $t$ is approximated by a series of steps of width $\triangle t$ and that the ordinate of the $m$-th step is 
$\omega_{m}^2$. With reference to our case, eq. (\ref{eq:lam}), let us suppose $f(t) > 0$. The transformation matrix 
$[M(m)]$, relating the displacement $x(m+1)$ and velocity $v(m+1)$ at the end of the interval $\triangle t_{m}$ to 
the respective values $x(m)$, $v(m)$ at the beginning, is obtained by solving the equation of motion, subject to 
the initial conditions $x = x(m)$ and $v = v(m)$. Since $f(t) = \omega_{m}^2$ is constant over the interval 
$\triangle t_{m}$, the equation of motion is that of a simple harmonic oscillator so that the following result is 
easily obtained: 
\[ M(m) = \left ( \begin{array}{ll}
\cos(\omega_{m} \triangle t)             &   \sin(\omega_{m} \triangle t)/\omega_{m} \\                                                                                                        \\
-\omega_{m} \sin(\omega_{m} \triangle t) &   \cos(\omega_{m} \triangle t)                                                                                                         \
         \end{array}
         \right ). \]

The complete transformation matrix $[M(T)]$, relating the initial and final column vectors, is obtained 
by taking the product of all the matrices for the entire time interval, i.e.:

\begin{equation}
M(t) = \prod M(m). 
\label{eq:prod}
\end{equation}

The question of stability is readily resolved by determining the eigenparameters of $M(T)$, that is to say, by 
solving a quadratic equation. 

\begin{figure}[htbp]
\centerline{\includegraphics[angle=-90,width=.8\textwidth]{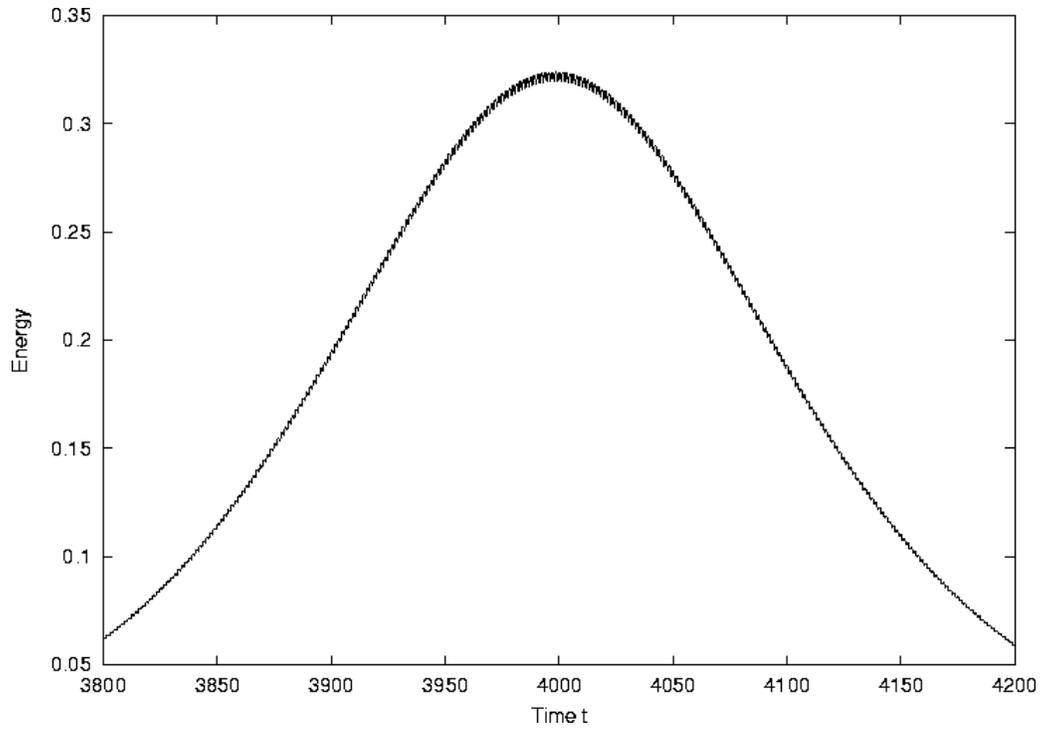}}
\caption{Energy of the mode 15 vs t for $\epsilon \mu = 0.005$.}
\label{fig:en15pi}
\end{figure}

\begin{figure}
\centerline{\includegraphics[angle=-90,width=.9\textwidth]{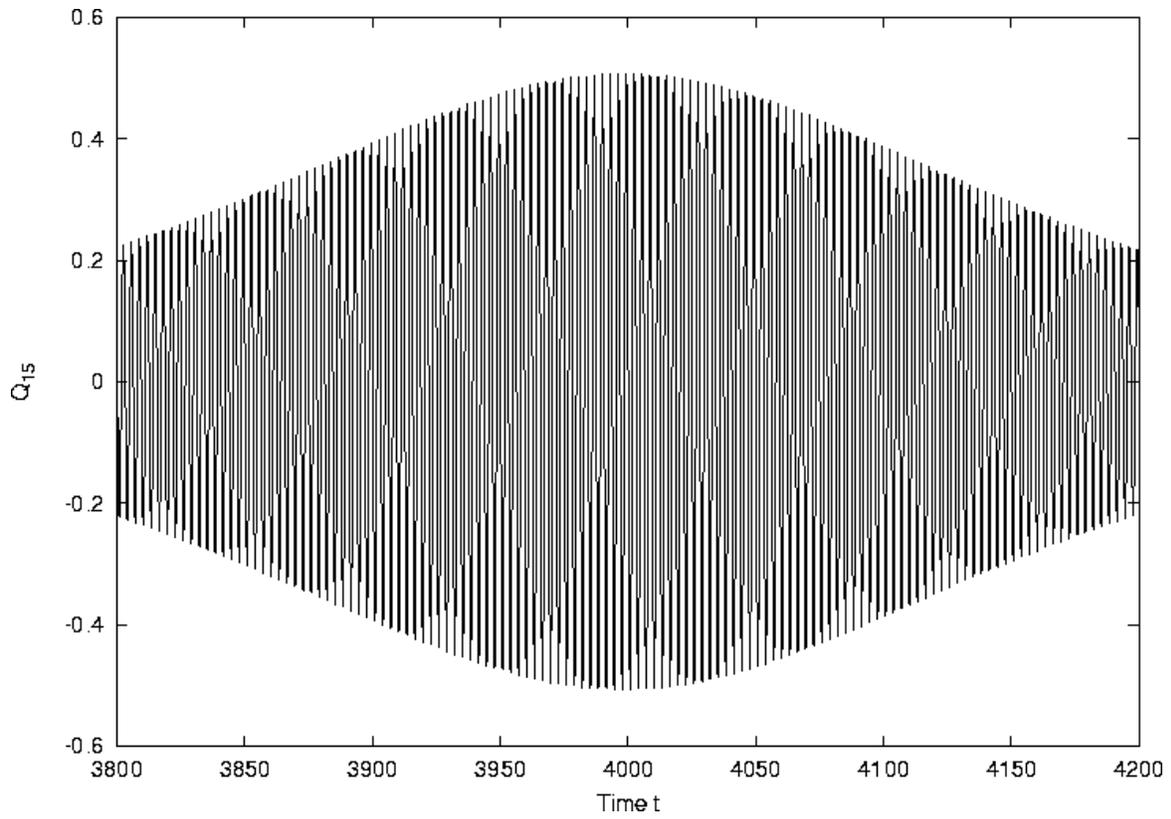}}
\caption{$Q_{15}$ vs t for $\epsilon \mu = 0.005$.}
\label{fig:q15pi}
\end{figure}

\begin{figure}[htbp]
\centerline{\includegraphics[angle=-90,width=.6\textwidth]{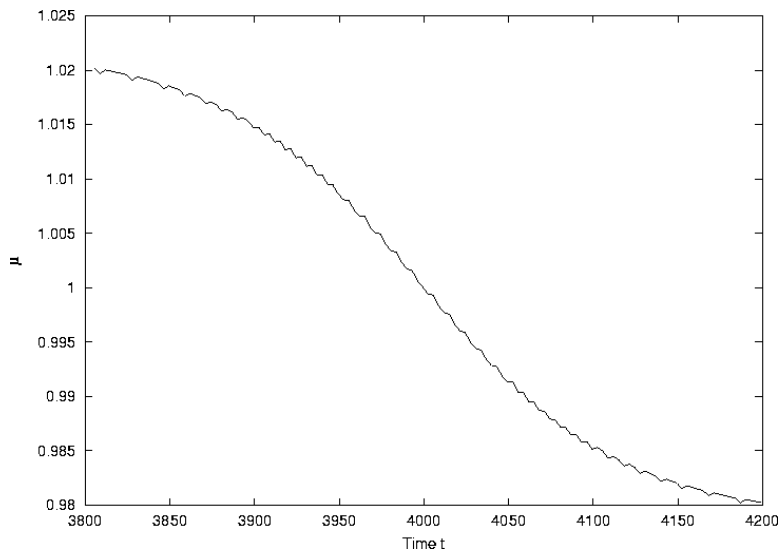}}
\caption{$\mu$ vs t for $Q_{15}$, $N = 32$ and $\epsilon \mu = 0.005$.}
\label{fig:mur}
\end{figure}

The  pulsed behaviour of the energy of the nonlinear one-mode solution and the consequent pulsed behaviour 
of the linear modes can then be attributed to the time behaviour of the parameters 
$\mu_{r}$, when $\beta > \beta_{t}$. Let us consider the Figs. \ref{fig:en15pi} and \ref{fig:q15pi} which 
show, for $N/2 = 16$ and $\beta = 0.005$, respectively the first energy pulse of the mode $15$ and the 
corrispondent time variation of $Q_{15}$. 
\eject
From numerical data we have verified that, in our case, namely in the case of eq. (\ref{eq:lam}), we have 
$X_{r}(t + T) = - X_{r}(t)$ and so $X_{r}(t + 2T) = X_{r}(t)$, where $T$ is the period of the Jacobian 
function $cn^{2}(\Omega_{N/2}t;k)$. For $\beta = 0.005$, from eqs. (\ref{eq:omen}) and (\ref{eq:kappa}), we have 
$\Omega_{N/2} \approx \omega_{N/2}$ and $cn^{2}(\Omega_{N/2}t;k) \approx \cos^{2} \omega_{N/2} t$. Since 
$\omega_{N/2} = 2$, the period of $\cos^{2}$ is $\pi/2$ and so $2 T = \pi$. From eq. (\ref{eq:per}), comparing the 
values of $Q_{15}$ each interval of $2 T = \pi$, we can obtain the value of the parameter $\mu_{r}$ which 
determines the stability or instability of the solution. The pulsed behaviour of the solution and of the energy 
can then be attributed to a time variation of $\mu_{r}$, due to the fact that, during the exchange of energy 
between the nonlinear one-mode and the other linear modes, eq. (\ref{eq:lam}) is no longer exactly valid. If we 
suppose that the one-mode solution $Q_{16}$ continues to oscillate with its normal frequency, but modulated in 
amplitude, during the energy exchange (see for example Fig. \ref{fi:q5}), then the parameter  $\mu_{r}$ varies during 
this exchange, causing the typical pulsed behaviour for $\beta > \beta_{t}$. Obviously, if $\beta$ is very large, 
many linear modes are involved, the exchange of energy is much greater and irreversible and the OMS does not 
recovers all its initial energy: the whole system tends asintotically to a state of energy equipartition.  
In figure \ref{fig:mur} the time behaviour of the parameter $\mu$ is shown for the solution $Q_{15}$, when the one-mode 
excited is the one-mode $Q_{N/2}$, for $N = 32$ and $\beta = 0.005$. With reference to Fig. \ref{fig:q15pi}, 
the values of $\mu$ are obtained calculating the ratio between two consecutive maxima of $Q_{15}$.       
It is clear from Fig. \ref{fig:q15pi} and Fig. \ref{fig:mur} the link between the exponential growth and the 
subsequent exponential decreasing of $Q_{15}$ with the variation of the parameter $\mu$ from values greater than 
one to smaller values.
\section{\bf Conclusions}
In this paper we have studied the problem of stability of the OMS's in the Fermi-Pasta-Ulam $\beta$ 
oscillator chain. Although this problem has been tackled many times in the last years, analytical results were 
available only for the case $N/2$, where $N$ is the number of particles in the chain. We have envisaged a simple 
numerical method, which tests the stability of solutions against the numerical errors introduced automatically by 
the numerical algorithm of integration of motion equations. The method reproduces the analytical results already 
known for case $N/2$, and allows to obtain the threshold energy density above which the OMS is unstable, in the 
other cases, namely the cases $N/4$, $N/3$, $\frac{2}{3}N$ and $\frac{3}{4}N$. 

We have found that, for each case, there is a characteristic energy density, above which, there is a large range 
of values of energy densities, before the chaotic region, in which the OMS presents an intermittent behaviour, in 
the sense that the nonlinear mode keeps its initial excitation energy for long times and periodically, abruptly, 
loses and recovers a fraction of this energy. We have verified that, for the case $N/2$, this characteristic 
energy density coincides, for large $N$, with the threshold energy density given by (\ref{eq:eps}). Then we have 
assumed this characteristic energy density as threshold energy density also in the other cases. 

We have also obtained that, varying $N$, the cases $\frac{N}{3}$, ~$\frac{2}{3}N$ and ~$\frac{N}{4}$, 
~$\frac{3}{4}N$ have the same threshold energy density. 

A tentative explanation of the intermittent behaviour, in  terms of Floquet's theorem for parametric oscillators, 
has been given.
\vfill\eject


\begin{thebibliography}{99}

\bibitem{fermi} E. Fermi, J.R. Pasta and S. Ulam, Collected papers
 of E. Fermi, ed. E. Segre (University of Chicago, Chicago,1965). 

\bibitem{ford} J. Ford, J. Math. Phys. {\bf 2}, 387 (1961); J. Ford and J. Waters, 
J. Math. Phys., {\bf 4}, 1293 (1963).

\bibitem{jac} E. A. Jackson, J. Math. Phys. {\bf 4}, 551 (1963).

\bibitem{nor} R. S. Northcote and R. B. Potts, J. Math. Phys. {\bf 5}, 383 (1964).

\bibitem{isr} F. M. Israiliev and B. V. Chirikov, Soviet Phys.-Doklady {\bf 11}, 30 (1966).

\bibitem{zas}  G. M. Zaslavsky and R. Z. Sagdeev, Soviet Phys.-JETP {\bf 25}, 718 (1967).

\bibitem{zab} N. J. Zabusky, {\it Nonlinear Partial Differential 
Equations}, Academic Press, 1967, p. 223.

\bibitem{poi} H. Poincar\'e, {\it Les Methodes Nouvelles de la 
M\'echanique Celeste}, Blanchard, Paris, 1987, Vol. {\bf 3}, p. 389.

\bibitem{fer} E. Fermi, Nuovo Cimento {\bf 25}, 267 (1923); {\bf 26} 105 (1923).

\bibitem{fo} J. Ford, Phys. Rep. {\bf 213}, 271 (1992).

\bibitem{kam} A. N. Kolmogorov, Dokl. Akad. Nauk SSSR {\bf 98}, 527 (1954);
 V. I. Arnold, Russ. Math. Surv. {\bf 18}, 9 (1963);  J. Moser, Nachr. Akad. Wiss. 
Goettingen Math.-Phys. K1.2 1, 1 (1962).

\bibitem{saito} N. Saito, N. Ooyama, Y. Aizawa and H. Hirooka, Suppl. of Prog. Theor. Phys. {\bf 45}, 209 (1970).

\bibitem{livi} R. Livi, M. Pettini, S. Ruffo, M. Sparpaglione, A. Vulpiani,
  Phys. Rev. A {\bf 31}, 1039 (1985).

\bibitem{pettini} M. Pettini and M. Landolfi, Phys. Rev. A {\bf 41}, 768 (1990).

\bibitem{casetti} L. Casetti, M. Cerruti-Sola, M. Pettini and E. G. D. Cohen, Phys. Rev. E {\bf 55}, 6566 (1997).

\bibitem{kantz} H. Kantz, Physica D {\bf 39}, 322 (1989).

\bibitem{vulpiani} R. Livi, M. Pettini, S. Ruffo, M. Sparpaglione, A. Vulpiani,
  Phys. Rev. A {\bf 28}, 3544 (1983).

\bibitem{poggi} P. Poggi and S. Ruffo, Physica D {\bf 103}, 251 (1997).

\bibitem{bud} N. Budinsky and T. Bountis, Physica D {\bf 8}, 445 (1983).
 
\bibitem{fla} S. Flach, Physica D {\bf 91}, 223 (1996).

\bibitem{cas} L. Casetti, Phys. Scripta {\bf 51}, 29 (1995).

\bibitem{landau} L. D. Landau and E. M. Lifshitz, {\it Mechanics}, Pergamon, London, 1960.

\bibitem{fame} N. Fameli, F. L. Curzon and S. Mikoshiba, Am. J. Phys. {\bf 67}, 127 (1999).

\bibitem{pipes} L. A. Pipes and L. R. Harvill, {\it Applied Mathematics for Engineers and Physicists}, 
Mc Graw - Hill, New York, 32nd ed., 1970.

\bibitem{yorke} E. D. Yorke, Am. J. Phys. {\bf 46}, 285 (1978).




\[
\]
\end{thebibliography}
\end{document}